\documentclass[10pt, preprint]{elsarticle}
\journal{arXiv.org}

\usepackage[usenames, dvipsnames]{xcolor}
\usepackage[a4paper, portrait, margin=1in]{geometry}
\usepackage{amssymb,amsmath}
\usepackage{graphicx, soul, overpic}
\usepackage{hyperref}
\usepackage{caption, subcaption}
\usepackage{tikz, tcolorbox}
\usepackage{comment,todonotes}

\newcommand{\lb}{\left(}
\newcommand{\rb}{\right)}
\newcommand{\beq}{\begin{equation}}
\newcommand{\eeq}{\end{equation}}
\newcommand{\un}[1]{\,\mathrm{#1}}
\vbadness=\maxdimen
\graphicspath{{./}{figs/}}

\newcommand{\rv}[1]{\textcolor{black}{#1}}
\newcommand{\rvs}[1]{\textcolor{black}{#1}} 
\patchcmd{\pprintMaketitle}
 {\hrule}
 {\clearpage\hrule}
 {}{}
\appto\endfrontmatter{\clearpage}

\begin{document}

\begin{frontmatter}

\title{Modelling and optimisation of water management in sloping coastal aquifers with seepage, extraction and recharge}
\vspace{2cm}
\author[a]{Raka~Mondal\fnref{equalcon}}
\author[a]{Graham~Benham\fnref{equalcon}}
\author[a]{Sourav~Mondal\fnref{sm}\corref{correspondingauthor}}
\author[b]{Paul~Christodoulides}
\author[c]{Natasa~Neokleous}
\author[d]{Katerina~Kaouri}
\address[a]{Mathematical Institute, University of Oxford, Oxford OX2 6GG, UK}
\address[b]{Faculty of Engineering and Technology, Cyprus University of Technology, 3036 Limassol, Cyprus}
\address[c]{Water Development Department, Limassol District Office, Cyprus}
\address[d]{School of Mathematics, Cardiff University, CF24 4AG, UK}

\fntext[equalcon]{These authors have contributed equally and are joint first authors.}
\fntext[sm]{Present address: Department of Chemical Engineering, Indian Institute of Technology Kharagpur, Kharagpur 721302, India}
\cortext[correspondingauthor]{Corresponding author, email: smondal@che.iitkgp.ac.in, Tel: +91-3222-304580}

\date{\today}

\begin{abstract}
We consider the management of sloping, long and thin, coastal aquifers. We first develop a simple mathematical model, based on Darcy flow for porous media, which gives the water table height, and the flow velocity as a function of the underground seepage rate, the recharge rates and the extraction rates, neglecting sea water intrusion. We then validate the model with recent data from the Germasogeia aquifer which caters for most of the water demand in the area of Limassol, Cyprus. The data is provided over a three-year period by the Cyprus Water Development Department (WDD), the governmental department managing the aquifer. Three different models of the recharge sources have been considered and it was found that Gaussian sources give the closest agreement with data. Furthermore, based on our model, we subsequently develop an optimised recharge strategy and identify the optimal recharge rates for a desired extracted water volume while the water table height is maintained at the acceptable level. We study several scenarios of practical interest and we find that we can achieve considerable water savings, compared to the current empirical strategy followed by WDD.  Additionally, we model the transport of pollutants in the aquifer in  the case of accidental leakage, using an advection-diffusion equation and the concentration is determined in the aquifer for an ongoing contamination and two pulsed contaminations (pulse duration 1 day and 30 days, respectively). We find that in the case of an undetected and unhindered contamination (worst case scenario) the aquifer would get polluted in about three years. Also, we find that double recharge rates flush the pollutant out of the aquifer faster.  Finally, to incorporate the possibility of sea water intrusion, which can render aquifers unusable, we develop a new, transient two-dimensional model of groundwater flow based on the Darcy-Brinkman equations, and determine the position of the water table and the seawater-freshwater interface for conditions of drought, moderate rainfall and flooding. \rvs{The validation of the new seawater intrusion modelling approach has been carried out via comparison with a widely-accepted code.}
\end{abstract}

\begin{keyword}
Aquifer; mathematical modelling; sea-water~intrusion; water~management; Darcy flow; Darcy-Brinkman equations
\end{keyword}

\end{frontmatter}

\section{Introduction}
Climate change, characterised by unpredictable seasonal fluctuations, a rise in annual temperatures and a growing population make natural water conservation one of the most important environmental challenges of the 21\textsuperscript{st} century, especially in locations with limited fresh water resources.

In island countries and coastal regions (for example, Cyprus \cite{theodossiou2004application}, Texas, USA \cite{wanakule1986optimal}, Turkey \cite{hallaji1996optimal}, Morocco \cite{sedki2011simulation}, Taiwan \cite{willis1988planning} and Greece \cite{mantoglou2003pumping}) where a significant part of the water supplies rely on aquifers, drought and over-extraction can lead to a drop in the aquifer water table. This drop increases the risk of sea water intrusion, which can render the aquifers unusable for a long period of time. In the case of drought, the water table lowers due to a decrease in the natural underground seepage flow in the aquifer. In order to maintain the water table and mitigate the risk of sea water intrusion, many coastal aquifers are replenished by pumping water back into the aquifer. Therefore, the water table height and consequently the management of aquifers requires a balance between the natural underground seepage, the extraction of water for consumption and the artificial recharge of water. 

By improving aquifer management strategies, it is possible to prevent the risk of sea water contamination and, hence, reduce the dependence on alternative desalination technologies, which are energy intensive, expensive and often not feasible for small economies or in the long term.

In the past three decades, there have been many studies (both laboratory experiments and mathematical modelling) on water management in coastal aquifers. The most notable mathematical strategy for modelling the aquifer water table is the Dupuit-Forchheimer approximation \cite{basha2005theoretical, castro2012steady, van2017interface, koussis2012analytical, koussis2015correction, wooding1966groundwater}, which neglects the pressure variation in the vertical direction and considers the aquifer to be long and thin. In this approximation the Darcy equations for flow in porous media are integrated across the aquifer depth to obtain a nonlinear partial differential equation for the water table height. However, in the studies mentioned above the interplay between seepage flow rates, recharge rates and extraction rates have not been studied and in only a few cases (e.g. \cite{koussis2012analytical, koussis2015correction}) was the developed model quantitatively compared to field data in the long term.

Moreover, there have been several reports on the controlled optimisation of the water management in the last three decades \cite{theodossiou2004application, wanakule1986optimal, hallaji1996optimal, sedki2011simulation, willis1988planning, mantoglou2003pumping, jones1987optimal, makinde1989optimal, cheng2000pumping, casola1986optimal}. 
Several optimisation techniques for different objectives were used. For example, Casola et al. \cite{casola1986optimal} performed a time-dependent optimal control of pumping costs,  Makinde et al. \cite{hallaji1996optimal} used feedback control of the water table height with respect to a critical level, and Shamir et al. \cite{shamir1984optimal} used a network-based linear programming approach for multi-objective optimisation. Some of these studies modelled sea water intrusion 
\cite{hallaji1996optimal, sedki2011simulation, willis1988planning, mantoglou2003pumping, cheng2000pumping} 
and some of them used field data from various locations, such as Cyprus \cite{theodossiou2004application}, Texas, USA \cite{wanakule1986optimal}, Turkey
\cite{hallaji1996optimal}, Morocco \cite{sedki2011simulation}, Taiwan \cite{willis1988planning} and Greece \cite{mantoglou2003pumping}.

The most comprehensive work on optimisation was by Mantoglou et al. \cite{mantoglou2003pumping}, and it is very relevant to the work we present here. The model in \cite{mantoglou2003pumping} for the water table height incorporates sea water intrusion using the single potential formulation of Strack \cite{strack1976single}, and a non-linear constrained optimisation method is carried out to find the optimum pumping rates for a confined, non-sloping aquifer. However, the optimisation protocol does not incorporate the balance between underground seepage, recharge and extraction and a point source/sink formulation is assumed for the recharge/extraction process respectively. Here we develop an optimisation protocol for sloping aquifers that also incorporates the balance between seepage, recharge, and extraction.
\rv{Unlike the study of Mantoglou et al. we do not incorporate a sea water intrusion model into our optimisation framework, but instead we impose a constraint that the water table must stay above a minimum critical level. This constraint serves as a proxy for avoiding sea water intrusion.}
\rv{Furthermore, considering that the recharge water disperses longitudinally as it enters the aquifer from the injection points at the surface, we model this with several realistic distribution functions, finding close agreement with the field data. There is nothing of probability or statistics being referred to here.
}

In the case of sea water intrusion, the two main modelling approaches can be divided into those in which the interface between sea water and fresh water is assumed to be sharp (i.e. no diffusion) and those in which the interface is considered diffuse. A simple and popular sharp-interface model, based on the assumption that the fresh water potential lines are vertical, is the Ghyben-Herzberg (GH) relation \cite{badon1889nota, herzberg1901wasserversorgung}. The single potential formulation of Strack \cite{strack1976single} is a sharp-interface model based on the GH relation (see also \cite{mantoglou2003pumping, koussis2015correction}). Strack uses the method of images to find an analytical solution, assuming the water is recharged and extracted at point sources and sinks, respectively. In \cite{de1981variational} a streamfunction-vortex formulation is used instead which models the sharp interface with a distribution of vortices \cite{van2017interface}. Other sharp-interface models use conformal mapping techniques in the hodograph plane \cite{charmonman1965solution, henry1959salt, rumer1968salt}. However, the above models, besides being complex, do not allow for more realistic distributions of recharge and extraction locations.

In diffuse-interface models, Henry \cite{henry1964effects} used a Boussinesq assumption and an advection-diffusion equation to model the concentration of salt in water in the aquifer. Many others have followed Henry's diffuse-interface model \cite{bolster2007analytical, croucher1995henry, henry1964effects, simpson2004improving, dentz2006variable} but these models, although more realistic than sharp-interface models, are much more computationally intensive, usually without a significant increase in accuracy.

Additional to the risk of sea water contamination, in aquifers near residential and agricultural areas there is also risk of contamination due to leaking of surface chemicals (e.g. pesticides, fertilisers, construction chemicals and human waste) \cite{bolster2007analytical, bear2010modeling}. However, since the aquifer water quality cannot be monitored everywhere and continuously, it is important to model and study contamination transport in the aquifer and be able to make predictions about the spread and rate of transport of the contaminants. In most cases hydrodynamic dispersion is quite slow so it is quite likely that by the time contamination is detected it has affected a large part of the aquifer, and this makes modelling even more essential. Models of contaminant transport also provide an estimate of the time required to flush the contaminant naturally, before it can be reused \cite{bear2010modeling,zheng2002applied}.

In the current study, we develop a model for the aquifer, based on the Dupuit-Forchheimer approximation, and we predict the \rv{underground} water table height as a function of the recharge rates \rv{(from injection at specific points)}, extraction rates and the natural underground seepage rate. \rv{The natural underground seepage originates from water entering the aquifer at the bottom of the upstream dam.}
In Section 2 the predicted water table is validated with field measurements of the Germasogeia aquifer over a three-year period\footnote{The current work emanated from the 125\textsuperscript{th} European Study Group with Industry (www.esgi-cy.org), a problem-solving workshop which took place in Cyprus, in December 2016.}. The aquifer, which leads to the Mediterranean Sea, has a small inclination angle, there is a dam upstream (the source of underground seepage) and it serves most of the demand in the town of Limassol, Cyprus. The aquifer is managed by the Water Development Department (WDD), the governmental department responsible for water management in Cyprus. We explore the balance between the extraction rates, the recharge rates, and the seepage rate, and how this balance affects the water table height in the aquifer. Initially, we model the recharge and extraction sources as point sources/sinks, respectively, and then extend our formulation to \rv{line sources, where the total discharged amount is the same as in through the point injection}.

In Section 3 we use a numerical optimisation method to identify the optimal recharge rates for given extraction and seepage rates. The objective of the optimisation is to minimise the total recharge whilst maintaining the water table above a critical level (so that no sea water intrusion takes place). The optimisation method not only allows for variation of the recharge rate, but also changing the number and location of the recharge sources. We can, thus, identify potential water savings, compared to the empirical protocol currently followed by WDD and make recommendations on how to improve the existing procedure by relocating the recharge sources or adding new ones.

Finally, in Section 4 we model sea water intrusion using the Darcy-Brinkman equations \cite{furman2008modeling, joodi2010development, chen2010asymptotic,neale1974practical} combined with a sharp-interface model based on the GH relation. We use this model to investigate how recharge, extraction and seepage affect sea water intrusion. 
\rv{Furthermore, the results of this approach are compared to those of a popular variable density groundwater tool - SUTRA, to establish reliability of the proposed model.}
\rvs{It may be emphasized that the present approach is computationally simpler, as we only solve for the momentum equations at the interface. The mass transport equations are not solved for, unlike the case with the Henry's approach. Generally, the available sharp interface codes are based on the physics of solving the species transport equation (creating density gradients) in the high-Peclet number case, using an optimized numerical algorithm or through the Ghyben--Herzberg approach (which inherently involves some critical assumptions) \cite{llopis2014discussion}. However, here we solve the fluid momentum equations for an interface, with the idea of a thin mass transfer boundary layer which is the case with high Peclet number, applicable for any practical scenario of coastal sea water intrusion. }
\rvs{Additionally, in the case of contamination of the aquifer, we predict the contaminant spread and speed in an aquifer using an advection-diffusion equation, where the water velocity is given by the model developed in Section 2. We also make appropriate recommendations for minimising the contamination impact. This is detailed in the Appendix C of the paper. }

\section{Modelling the water table height in a sloping, long and thin, coastal aquifer}\label{dupuit-forch}

We consider a gravity-driven flow in a sloping, porous aquifer. We assume that the aquifer is wide enough so that width effects are negligible and the problem can be approximated in two dimensions. This is a good approximation for the Germasogeia aquifer, in the district of Limassol, Cyprus. 
\rv{For other aquifers, where width effects are more prominent, a full three-dimensional model may be more suitable. }

\begin{figure}
\centering
\footnotesize
\begin{overpic}[width=0.5\textwidth]{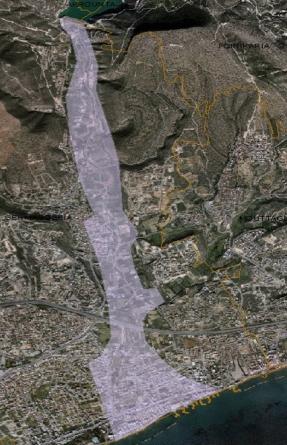}
\end{overpic}
\caption{Aerial view of the Germasogeia aquifer (source: Cyprus Water Development Department)}
\label{fig:GermasogeiaAquifer}
\end{figure}

\begin{figure}
\centering
\begin{tikzpicture}[scale=0.6]
\draw[<->,line width=2] (0,0) -- (0,3); 
\draw[<->,line width=2] (0.1,-0.1) -- (14.9,-0.1); 
\draw[<->,line width=2] (15.5,0) -- (16,3); 
\draw[line width=2] (0,3) -- (15,0); 
\draw[line width=2] (0,3) -- (0.6,6);
\draw[->, line width=2] (0,3) -- (0.4,5);
\draw[->, line width=2] (0,3) -- (2,2.6);
 \node at (-0.2,5) {\large $z$};
\node at (2.3,3.0) {\large $x$};
\draw[line width=2] (15,0) -- (15.6,3);
\draw[line width=2] (0.6,6) -- (15.6,3); 
\node at (7,-0.7) {\large $L$};
\node at (-0.7,1.5) {\large $H$};
\node at (16.5,1.5) {\large $H_b$};
\draw [blue,dashed](0.2,4) parabola bend (10,3) (20,3);
\node at (18.5,1.5) { Sea};
\node at (5,1.5) {Bedrock};
\draw[->,line width=2] (5,4) -- (5.5,6); 
\draw[<-,line width=2] (8,3.5) -- (8.5,5.5); 
\node at (5,6.5) {Extraction (Sink)};
\node at (10.5,6) {Recharge (Source)};
\node at (12,4.5) {Ground surface};
\draw[->,line width=1] (-0.5,3.5) -- (1,3.2);
\node at (-0.9,3.5) {\large $Q$};
\node at (-1.3,5) {Dam};
\draw [line width =2,gray] (12,-0.1) arc [radius=1, start angle=180, end angle= 140];
\node at (11,0.3) {\large $\alpha$};
\node at (10,2.5) {Water table: $z=h(x,t)$};
\end{tikzpicture}
\caption{Schematic diagram of the long and thin sloping aquifer related to the model equations \eqref{darcy1}--\eqref{darcy3}. The water table level is indicated with the blue dashed curve. The coordinate system $x$ and $z$ is taken respectively along and perpendicular to the bedrock, which is assumed to be flat and $\alpha$ is the angle to the horizontal level.}
\label{fig:schem}
\end{figure}
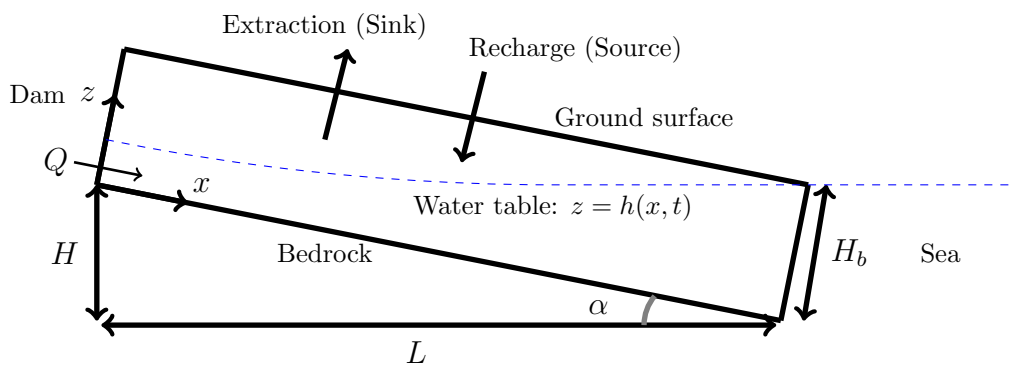

This alluvial aquifer lies in the Germasogeia river valley and extends from the Germasogeia dam to the coast. It is $5.5$km long, and its depth ranges from 35m near the dam to about 55m near the sea while its width ranges from 100m to 800m (wider close to the sea)--see Fig.~\ref{fig:GermasogeiaAquifer}. The WDD (Limassol district) manage this aquifer and the groundwater, which is naturally filtered in the aquifer, is extracted through boreholes at several locations and used to fulfill water needs in the greater Limassol area.  The WDD also artificially recharge the aquifer at four locations in order to maintain the water table at an acceptable level. Beneath the aquifer there is an impermeable bedrock, which is assumed to be flat, and the top boundary is the ground surface (see Fig.~\ref{fig:schem}). There is a saturated water table level at some height below ground level. Below the water table the aquifer is fully saturated and  approximately dry above it. Upstream \rv{of} the aquifer a dam holds a large body of water behind a concrete barrier, and there is water seepage through the barrier with volumetric flow rate $Q$. Rainfall contribution is accounted in the model through the dam seepage. Downstream, the water table meets the sea level at height $H_b$ above the bedrock. 

We choose a rotated coordinate system such that the $x$ direction is parallel to the bedrock level, inclined at a constant angle $\alpha$ to the horizontal, and the $z$ direction is perpendicular to the bedrock. $L$ and $H$ are the length and elevation of the aquifer, respectively, and $\tan \alpha = H/L$ (see Fig.~\ref{fig:schem}).
We denote by $(u,w)$ the velocity components in the $(x,z)$ directions, and by $p$ the pressure. The incompressible flow of a Newtonian fluid in a porous medium is governed by the continuity equation and the Darcy equations as follows:

\begin{align}
\label{darcy1}
\rvs{\frac{\partial u}{\partial x}+\frac{\partial w}{\partial z}}&=\rvs{s(x,t),}\\
\label{darcy2}
u&=-\frac{k(x)}{\phi \mu}\lb  \frac{\partial p}{\partial x}-\rho g \sin \alpha\rb,\\
w&=-\frac{k(x)}{\phi \mu}\lb  \frac{\partial p}{\partial z}+\rho g \cos \alpha\rb,\label{darcy3}
\end{align}
where $\rho$ is the density of the fluid (fresh water), $g$ is the gravitational acceleration constant, $\mu$ is the viscosity of the fluid and $\phi$ is the aquifer porosity, which we assume to be constants. We assume that the permeability of the porous medium $k(x)$ varies along the length of the aquifer. The rate of extraction (sinks) and recharge (sources) are accounted in $s(x, t)$ in the right hand side of \eqref{darcy1}. Later on, we discuss the form of $s(x,t)$ and different possible recharge and extraction distributions. The water table is denoted by $z=h(x,t)$ (blue dashed curve in Fig.~\ref{fig:schem}). The kinematic and dynamic boundary conditions are

\begin{align}
w=\frac{\partial h}{\partial t}+u\frac{\partial h}{\partial x} \quad& \mathrm{on} \quad z=h,\label{kin}\\
p=p_a \quad& \mathrm{on} \quad z=h,\label{dyn}
\end{align}
where $p_a$ is the atmospheric pressure. At the bedrock, we have the impermeability condition
\begin{equation}
w=0 \quad \mathrm{on} \quad z=0 \label{imp}.
\end{equation}
The dam upstream the aquifer is a source of underground seepage and we have the following boundary condition at the dam-aquifer barrier:
\begin{equation}
\int_0^{h} u\,dz=Q(t)/\ell\quad\mathrm{at}\quad x=0,\label{qbc}
\end{equation}
where $\ell$ is the typical dam width. Downstream of the aquifer the water table is at sea level
\begin{equation}
h=H_b \quad\mathrm{at}\quad x=L/\cos \alpha,\label{hbc}
\end{equation}
which we assume constant (ignoring tidal fluctuations). We non-dimensionalise all variables using the scalings
\begin{equation}
\begin{split}
&x=L\hat{x},\quad z=H\hat{z},\qquad t=\frac{ L^2}{K H}\hat{t},\qquad p=p_a+\rho g H\hat{p},\qquad u=\frac{K H}{ L}\hat{u},\\ &\quad w=\frac{K H^2}{ L^2}\hat{w},\qquad s=\frac{K H}{ L^2}\hat{s},\qquad h=H\hat{h},\qquad k=\kappa \hat{k},
\end{split}
\end{equation}
where $K=\kappa \rho g / \mu$ is the hydraulic conductivity and $\kappa$ is a typical permeability value. The inclination angle $\alpha$ is considered small ($\alpha\approx 0.01$ radians for the Germasogeia aquifer)  and the aquifer is assumed long and thin such that the aspect ratio $\epsilon=\tan\alpha$ is approximately
%
%
\begin{equation}
\epsilon  \approx \alpha.\label{small}
\end{equation}
Considering (\ref{small}), equations \eqref{darcy1}$-$\eqref{darcy3} become

\begin{align}
\rvs{\frac{\partial \hat{u}}{\partial \hat{x}}+\frac{\partial \hat{w}}{\partial \hat{z}}}&=\rvs{\hat{s},}\label{darcy1a}\\
\hat{u}&=\frac{\hat{k}}{\phi}\lb1- \frac{\partial \hat{p}}{\partial \hat{x}}\rb,\label{darcy2a}\\
\epsilon^2 \hat{w}&= -\frac{\hat{k}}{\phi}\lb 1+\frac{\partial \hat{p}}{\partial \hat{z}}\rb.\label{darcy3a}
\end{align}
If we ignore terms which are of order $\mathcal{O}(\epsilon)$ and smaller, equation \eqref{darcy3a} is reduced to the classical Dupuit-Forchheimer approximation \cite{basha2005theoretical, castro2012steady, van2017interface, koussis2012analytical, koussis2015correction, wooding1966groundwater}, which we integrate (using (\ref{dyn})) to get
\beq
\hat{p}=\hat{h}-\hat{z}.\label{press_int}
\eeq
Inserting \eqref{press_int} into \eqref{darcy2a} we find the horizontal velocity
\begin{equation}
\hat{u}=\frac{\hat{k}}{\phi}\lb 1-\frac{\partial \hat{h}}{\partial \hat{x}} \rb.\label{sea}
\end{equation}
Integrating equation \eqref{darcy1a} across the water table $0 \leq \hat{z} \leq \hat{h}$ and using the boundary conditions \eqref{kin}, \eqref{dyn} and \eqref{imp} (and the Leibniz integral rule), we get 
\rvs{
\begin{equation}
\phi \frac{\partial \hat{h}}{\partial \hat{t}} + \frac{\partial }{\partial \hat{x}}\lb \hat{k}\hat{h}\lb 1-\frac{\partial \hat{h}}{\partial \hat{x}}\rb\rb=\phi \hat{s} \hat{h}.\label{govh}
\end{equation}}
The boundary conditions \eqref{qbc} and \eqref{hbc} become

\begin{align}
\hat{h}\hat{k}\lb 1-\frac{\partial \hat{h}}{\partial \hat{x}}\rb=\hat{q}:\quad \mathrm{at}\quad \hat{x}=0,\label{nondimbc1}\\
\hat{h}=\hat{H}:\quad \mathrm{at}\quad \hat{x}=1,\label{nondimbc2}
\end{align}
where $\hat{q}(\hat{t})=\phi Q L/(K H^2 \ell)$ and $\hat{H}=H_b/H$. 
Equation \eqref{govh}$-$\eqref{nondimbc2} and an appropriate initial condition, can be solved for the water table height, assuming knowledge of the source/sink terms $\hat{s}$, the seepage rate $\hat{q}$ and the permeability $\hat{k}$. We represent the contribution to $\hat{s}$ from recharge sources by $\hat{R}$ and extraction sinks by $\hat{E}$, such that
\beq
\hat{s}=\hat{R}-\hat{E}.
\eeq
Below, we will discuss various distributions for $\hat{R}$ and $\hat{E}$, and compare them when validating the model with field data from the Germasogeia aquifer, over a three-year period (Oct 2013--Nov 2016). Currently, there are four recharge locations and nineteen extraction points in operation, distributed over the length of the aquifer. The dam seepage rate varies seasonally over this period (it depends on the overground water level). The extraction points are monitored every fortnight and the recharge rate is altered by WDD every fortnight depending on the demographic demand (inferred from the extraction rates). The time series data of the dam seepage, extraction and recharge rates is presented in \ref{appA} (Fig.~\ref{fig:timeseries_data}). The location of the four recharge and nineteen extraction points are given in the figure legends (Fig.~\ref{fig:timeseries_data}). The density and viscosity of fresh water are $\rho = 998~ \un{kg/m^3}$ and $\mu = 0.9\times10^{-3}~\mathrm{Pa.s}$, respectively. The porosity of the aquifer soil material is $\phi = 0.4$, as reported by WDD. The permeability of the soil is spatially varying, but by not more than $\pm 20\%$ of its mean value, which is given in terms of the conductivity as $K = 1.5 \times 10^{-3}~\un{m/s} = 130~\un{m/day}$. In our calculations, we have considered both a constant conductivity (mean value), and a spatially varying conductivity and found only minor differences in the predicted water levels. Therefore we proceeded with the assumption of constant conductivity (see Fig.~\ref{fig:permeability}(a)).

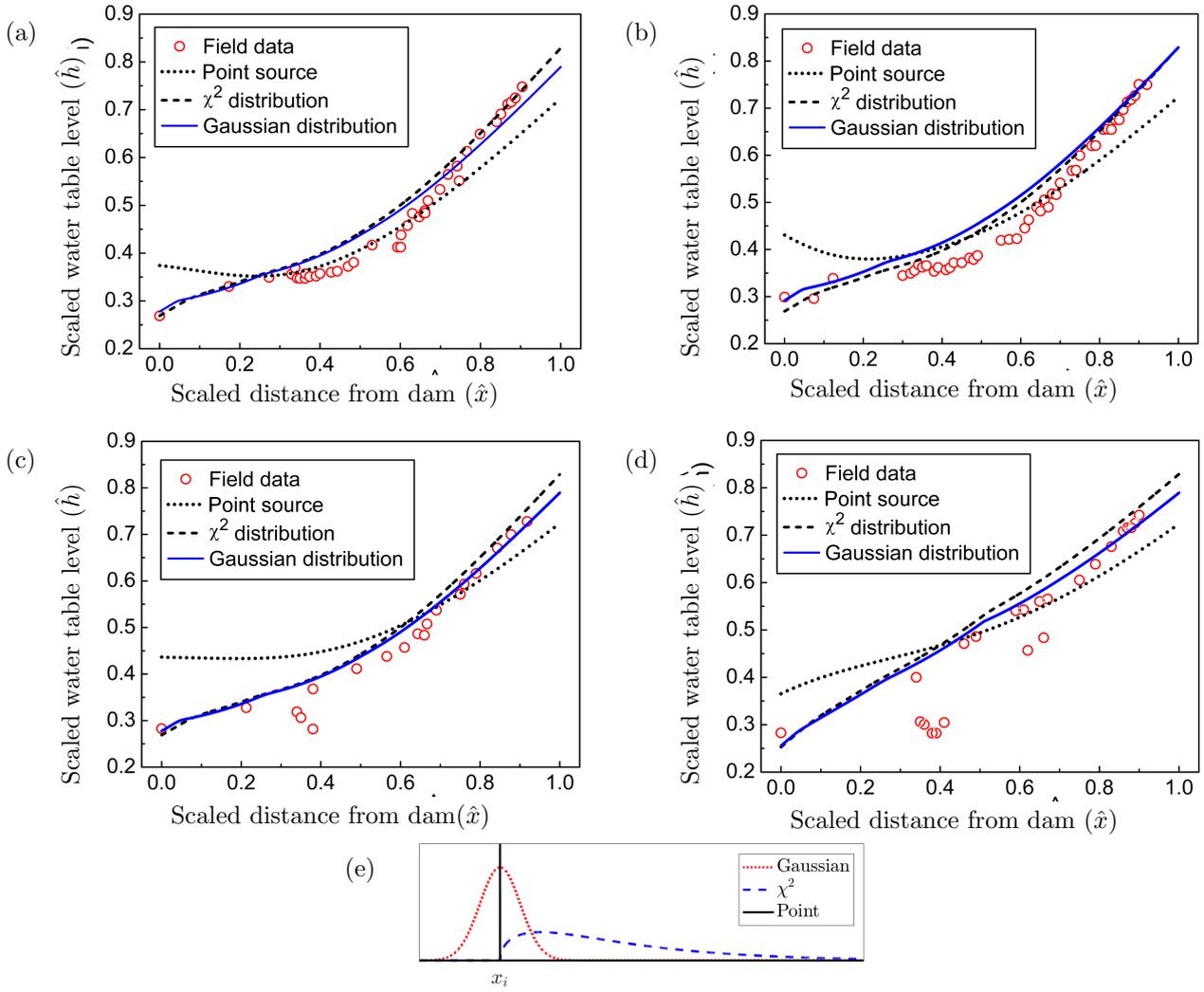
\begin{figure}[htbp!]
\centering
\begin{overpic}[width=0.46\textwidth, trim={0 0 0 0},clip]{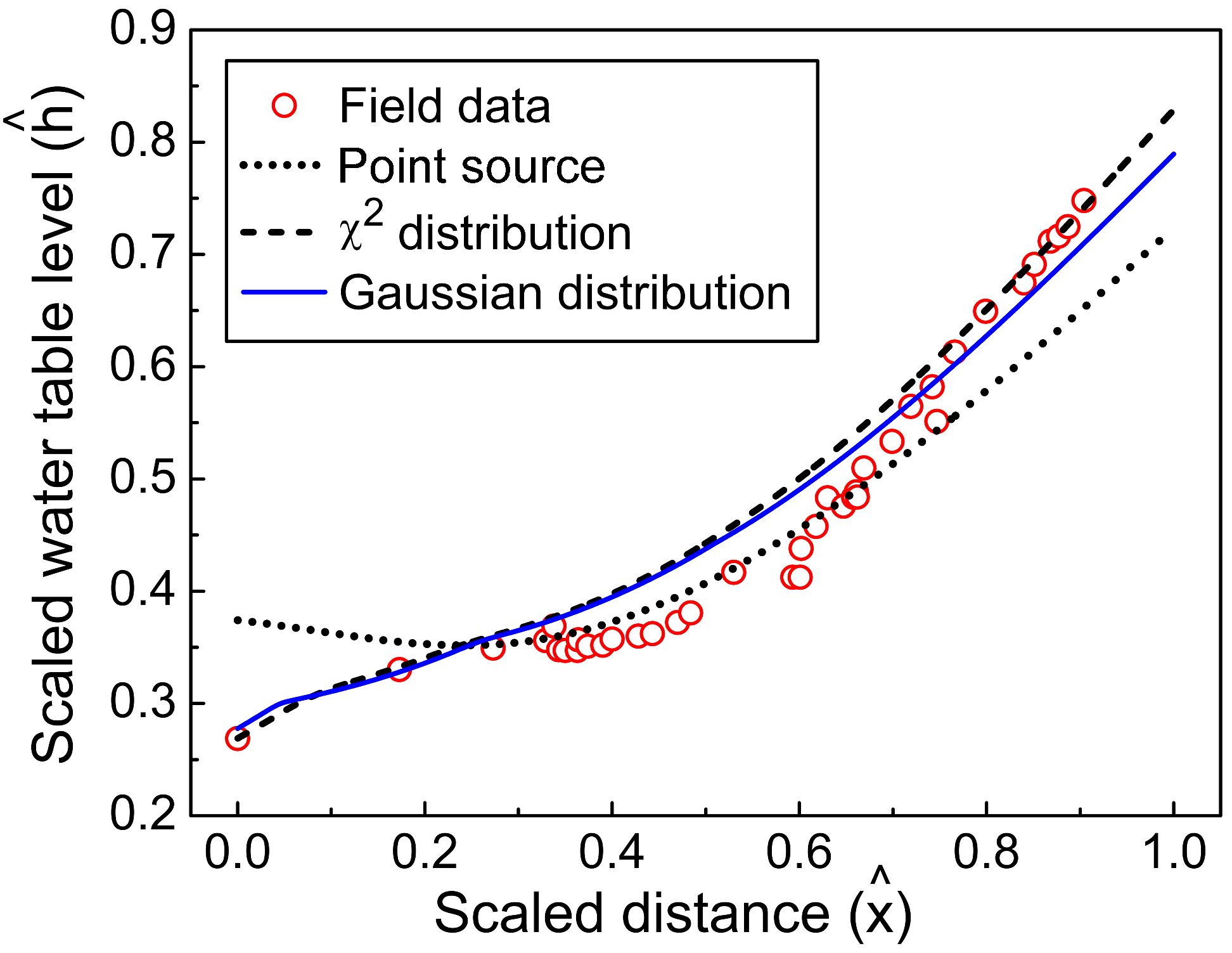}
\put(-10,70){(a)}
\put(-2,8){\rotatebox{90}{\colorbox{white}{Scaled water table level ($\hat{h}$)}}}
\put(20,1){\rotatebox{0}{\colorbox{white}{Scaled distance from dam ($\hat{x}$)}}}
\end{overpic}
\hfill
\begin{overpic}[width=0.46\textwidth, trim={0 0 0 0},clip]{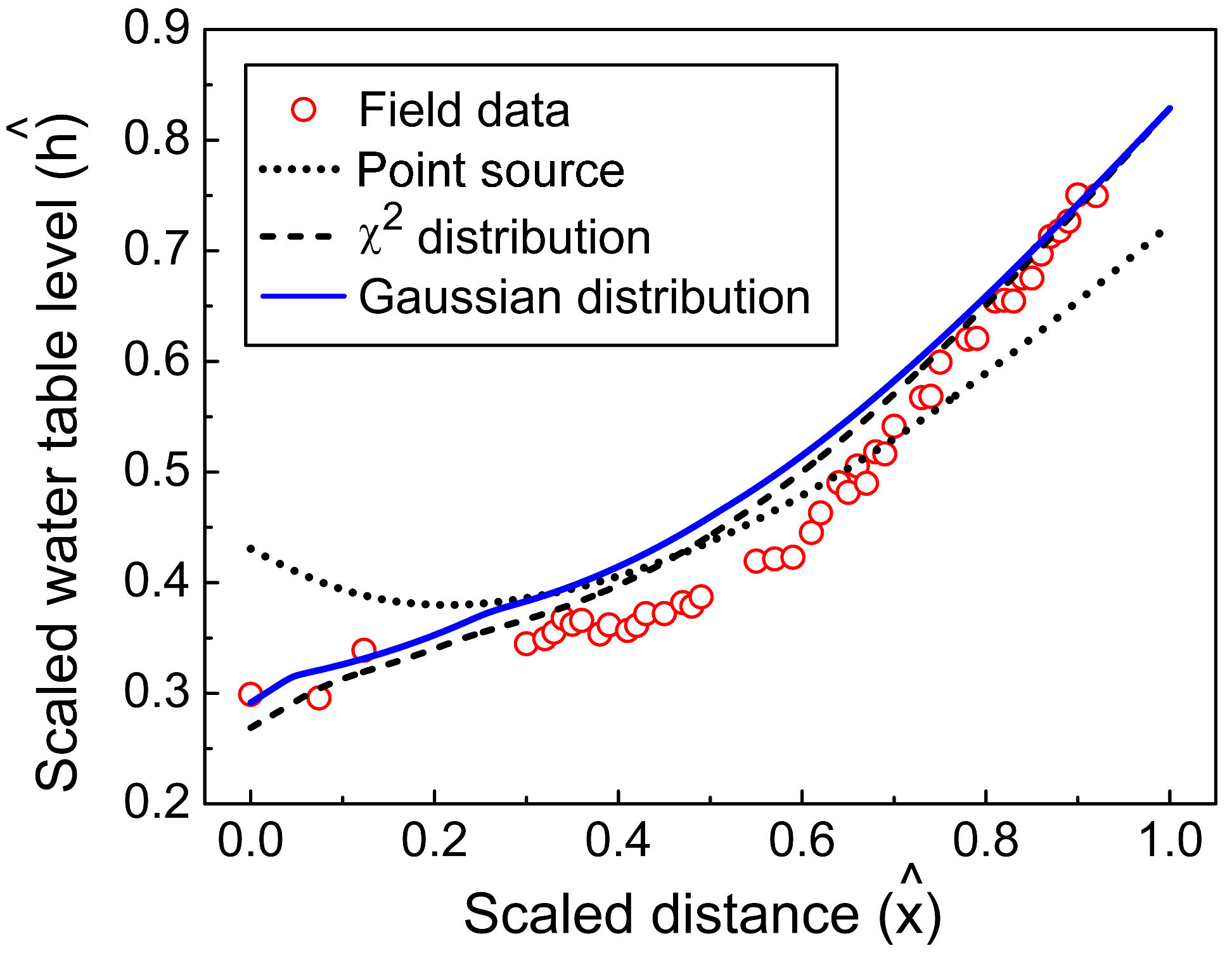}
\put(-10,70){(b)}
\put(-2,8){\rotatebox{90}{\colorbox{white}{Scaled water table level ($\hat{h}$)}}}
\put(20,2){\rotatebox{0}{\colorbox{white}{Scaled distance from dam ($\hat{x}$)}}}
\end{overpic}

\vspace{3mm}

\begin{overpic}[width=0.46\textwidth, trim={0 0 0 0},clip]{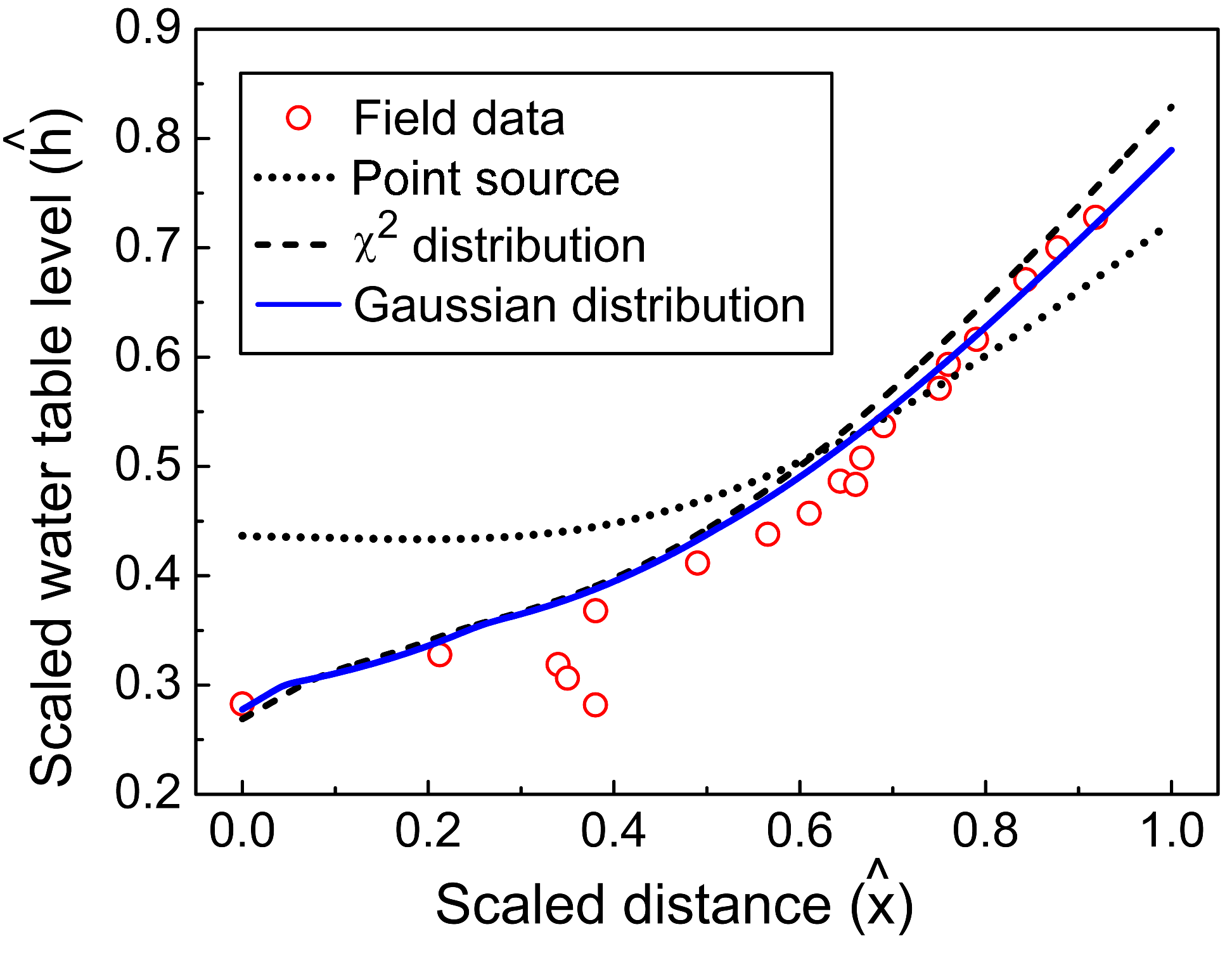}
\put(-10,70){(c)}
\put(-2,8){\rotatebox{90}{\colorbox{white}{Scaled water table level ($\hat{h}$)}}}
\put(20,2){\rotatebox{0}{\colorbox{white}{Scaled distance from dam($\hat{x}$)}}}
\end{overpic}
\hfill
\begin{overpic}[width=0.46\textwidth, trim={0 0 0 0},clip]{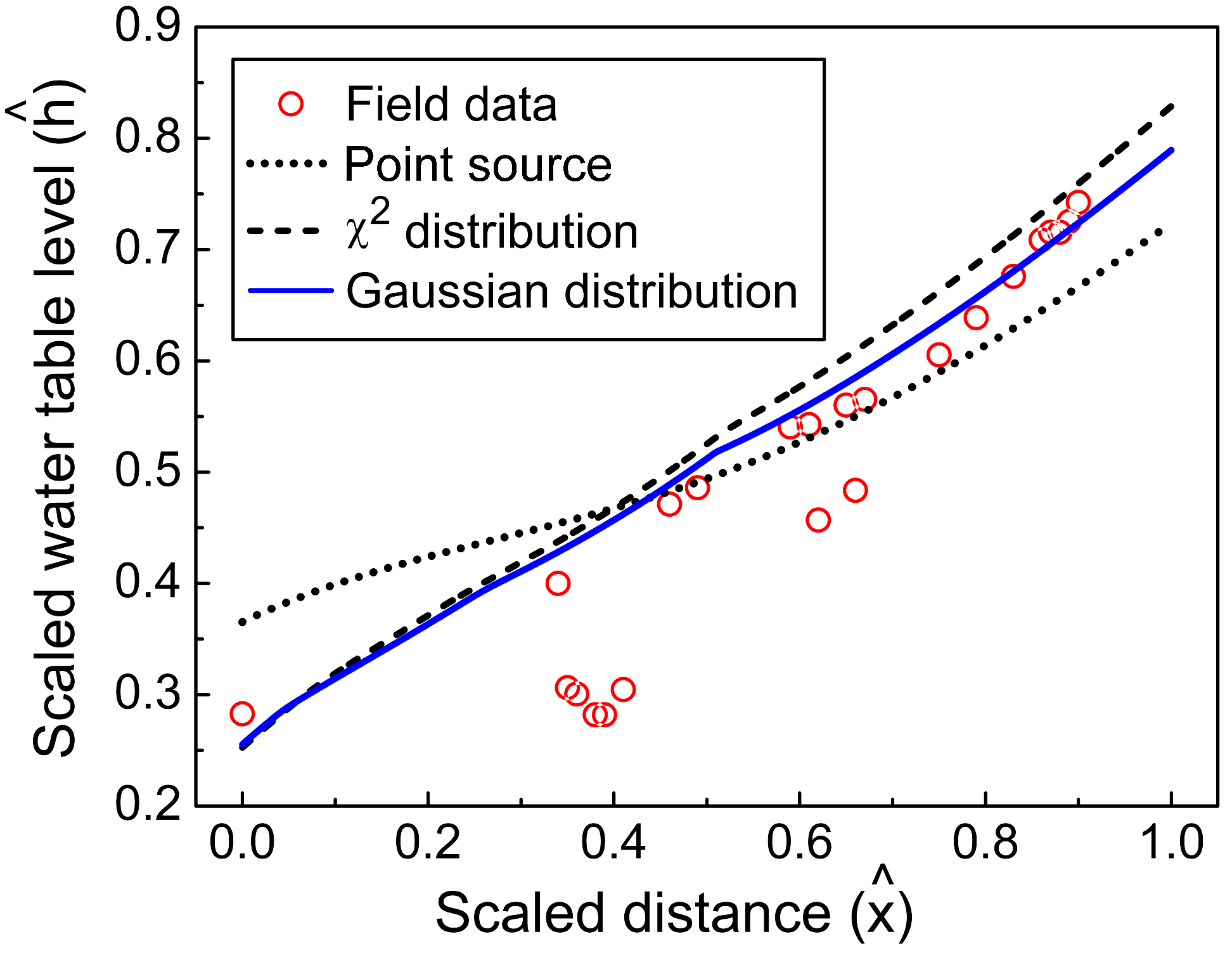}
\put(-10,70){(d)}
\put(-2,8){\rotatebox{90}{\colorbox{white}{Scaled water table level ($\hat{h}$)}}}
\put(20,1.2){\rotatebox{0}{\colorbox{white}{Scaled distance from dam ($\hat{x}$)}}}
\end{overpic}
\begin{overpic}[width=0.5\textwidth]{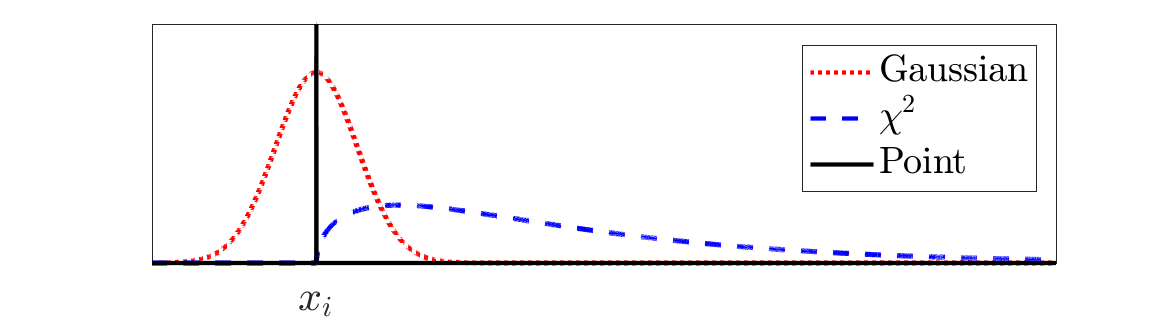}
\put(0,20){(e)}
\end{overpic}
\caption{Comparison of observed and simulated water table profiles (in the rotated frame), using different distributions for the recharge and constant permeability (130 m/day). Results are shown at the (a) $285^{th}$ day, (b) $570^{th}$ day, (c) $855^{th}$ day and (d) $1140^{th}$ day from the first measurement.  For the Gaussian distributions, we use $\sigma_1 = 0.01$,  $\sigma_2 = 0.02$,  $\sigma_3 = 0.001$, and  $\sigma_4 = 0.01$ which fit the data well. For the $\chi^2$ distribution, we found that $k=3$ is a good fit for all recharge points. The dam seepage, recharge and extraction rates used in the model correspond to the measured values throughout the period Oct 2013--Nov 2016, for the Germasogeia aquifer. \rv{(e) Illustration of the different distributions considered.}
}
\label{fig:matching}
\end{figure}

\begin{figure}[htbp!]
\centering
\footnotesize
\begin{overpic}[width=0.47\textwidth]{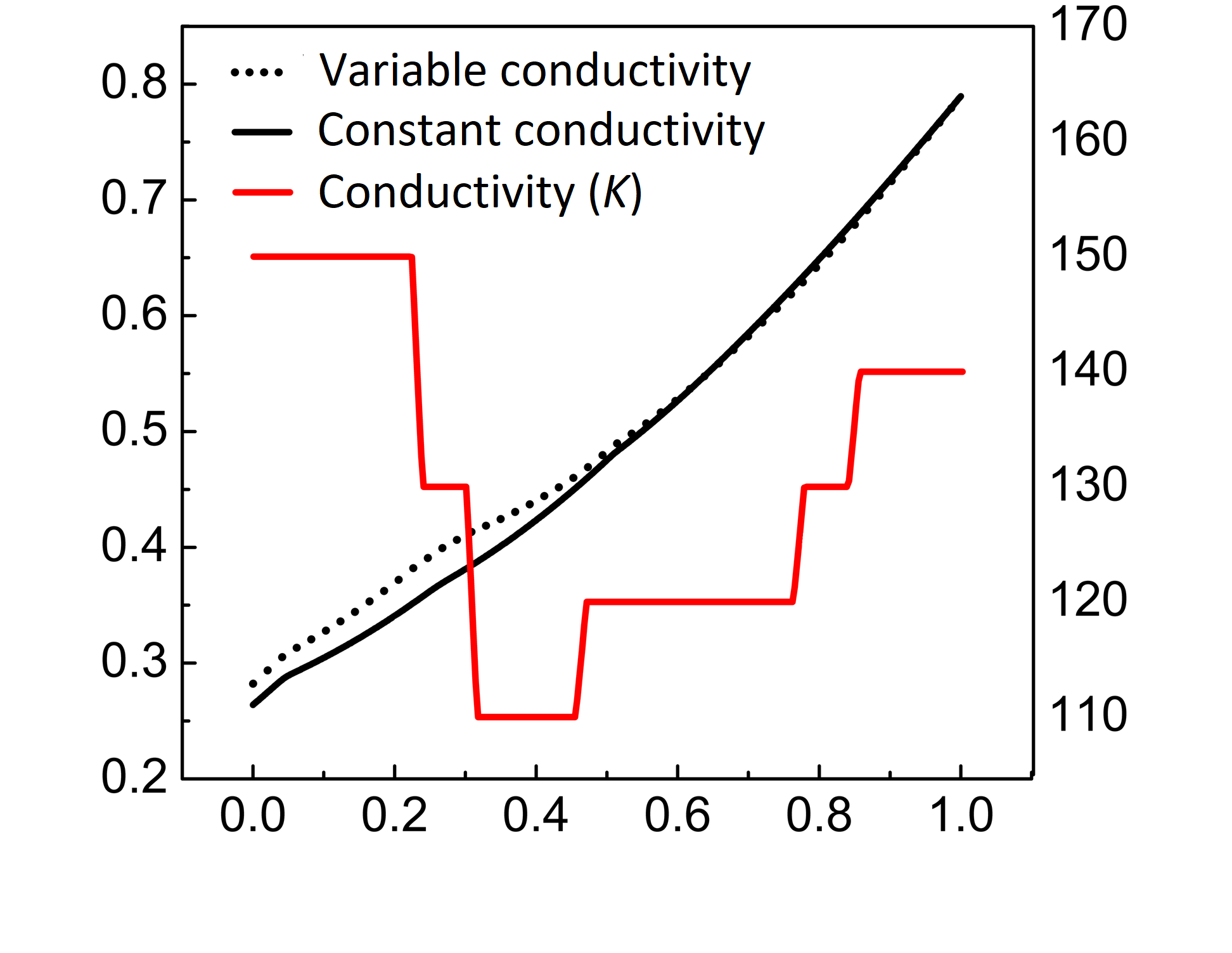}
\put(-10,70){\normalsize (a)}
\put(-2,12){\rotatebox{90}{\colorbox{white}{Scaled water table level ($\hat{h}$)}}}
\put(93,12){\rotatebox{90}{\colorbox{white}{Hydraulic conductivity ($m/d$)}}}
\put(18,4){\rotatebox{0}{\colorbox{white}{Scaled distance from dam ($\hat{x}$)}}}
\put(47,25){\vector(1,0){20}}
\put(60,50){\vector(-1,0){20}}
\put(32,32){\vector(-1,0){10}}
\end{overpic}
\hfill
\raisebox{5mm}{
\begin{overpic}[width=0.45\textwidth]{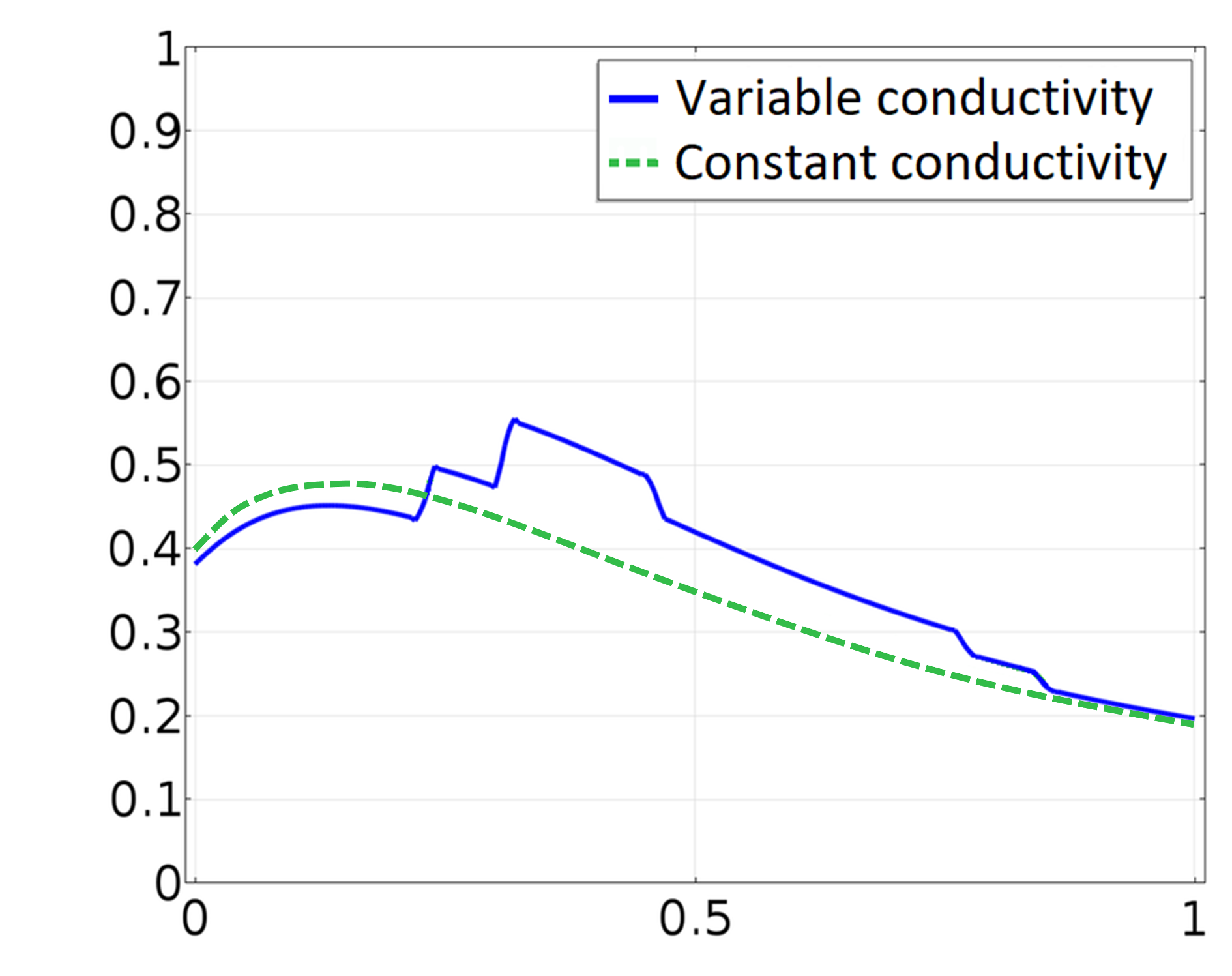}
\put(-10,75){\normalsize (b)}
\put(-2,23){\rotatebox{90}{\colorbox{white}{Scaled velocity ($\hat{u}$)}}}
\put(25,-5){\rotatebox{0}{\colorbox{white}{Scaled distance from dam ($\hat{x}$)}}}
\end{overpic}
}
\caption{Comparison of simulated (a) height profile $\hat{h}$ and (b) velocity $\hat{u}$ in the rotated frame, using both measured permeability (or conductivity) data and the constant spatially averaged value (which corresponds to a conductivity value of $K=130\,\mathrm{m/day}$). These results are after a month from the first observation data (Oct 2013).
}
\label{fig:permeability}
\end{figure}
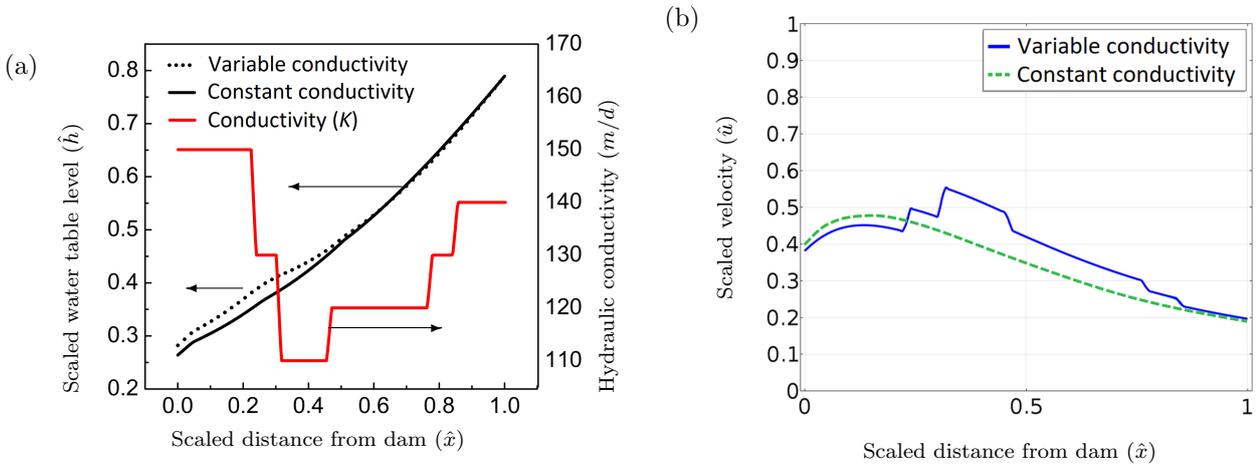

The recharge sources are initially considered to be point sources, on the ground. After the recharge water enters the porous medium, it percolates underground before reaching the water table. Since there are no observations of the percolation, in our model we incorporate this effect by considering three different recharge distribution patterns, \rv{sourced into the water table}. 
We consider point sources, symmetric Gaussian distributions and asymmetric $\chi^2$ distributions \rv{of the water profile fed into the water table.}
In the case of the point source distribution,
\begin{equation} \label{eq:point_source}
\hat{R} = \sum_{i=1}^{4} \hat{R}_i(\hat{t}) \delta \left( \hat{x}-\hat{x}_i \right).
\end{equation}
In the case of the Gaussian distribution,
\begin{equation} \label{eq:gauss_distr}
\hat{R} = \sum_{i=1}^{4} \hat{R}_i(\hat{t}) \frac{1}{\sqrt{2\pi \sigma_i^2}} \exp \left[ -\frac{(\hat{x}-\hat{x}_i)^2}{2\sigma_i^2}\right],
\end{equation}
where $\sigma_i$, \rv{the standard deviation, is the parameter which controls the spatial width of the flow rate distribution, fed into the aquifer water table from the surface at single point injections (recharge)}. Finally, in the case of the $\chi^2$ distribution
\begin{equation} \label{eq:chi_sq_distr}
\hat{R} = \sum_{i=1}^{4} \hat{R}_i(\hat{t}) \frac{1}{2^{k/2} \Gamma(k/2)} \left( \hat{x}-\hat{x}_i \right)^{\frac{k}{2} - 1} \exp \left[ -\frac{(\hat{x}-\hat{x}_i)}{2} \right],
\end{equation}
where $k$ is the degree of freedom of the $\chi^2$ distribution.
Unlike the recharge process, extraction is performed using pumping wells which are drilled deep underground beneath the water table. Therefore, their effect is not distributed and we can model extraction sinks with points, as follows
\begin{equation} \label{eq:point_extract}
\hat{E} = \sum_{i=1}^{19} \hat{E}_i (\hat{t}) \delta \left( \hat{x}-\hat{x}_i \right),
\end{equation}
where the values of $\hat{E}_i$ are taken from the Germasogeia field data. 
\rv{The different distributions considered in this study are illustrated qualitatively in Fig. \ref{fig:matching}(e).}
The initial condition for $\hat{h}$ is chosen according to the first measurement of the water table height (Oct 2013). Since the field measurements are obtained at discrete points throughout the aquifer, we have fitted a fourth order polynomial to generate a smooth curve for the initial table height. 

In Fig.~\ref{fig:matching} we display the measured water table height and the result of our model for the three different \rv{water distribution functions}. 
\rv{Here the $\sigma_i$ for the Gaussian distribution and the parameter $k$ in the $\chi^2$-distribution are fitting parameters for the water table field data. For the $\sigma_i$ the values are chosen using a least-squares error (residual) minimization for the field data values and the observed model results for all the observation times recorded in the field study. The objective function in the error minimization is 
\begin{equation}
Err = \sum_{T_N} \sum_{D_N} \left[ \frac{\hat{h} - \hat{h}_f}{\hat{h}_f} \right]^2
\end{equation}
where $\hat{h}_f$ represents the observed water table height, $D_N$ is the number of field measurement data points and $T_N$ is the number of time points in the three year period of Oct 2013 - Nov 2016. The values of the optimisation variables, which are the $\sigma_i$ in this case, are constrained within the range of 0.0002 to 0.1.  We have found $\sigma_1 = 0.01$, $\sigma_2 = 0.02$, $\sigma_3 = 0.001$, and  $\sigma_4 = 0.01$ using this minimization routine. 
The value of $k$ in the $\chi^2$ distribution is obtained from a similar least-squares error minimization. However, in this case only integer values constrained in the range of 3 to 6 are allowed for $k$. We have found $k=3$ to be the optimum value in this case.
There is also another reason behind the preference for choice of a $\chi^2$ distribution, as it asymmetric in nature and therefore realistically connects to the water flow pattern in the sloping aquifer (since water cannot travel uphill).
}

The comparison between the model and the data for the first 200 days is good since this data is close in time to the initial polynomial fit. Therefore, we only plot later comparisons at $285$, $570$, $855$ and $1140$ days after the initial time (Fig.~\ref{fig:matching}(a)-(d), respectively). It is evident from Fig.~\ref{fig:matching} that considering the recharge sources as point sources correlates poorly with the data, compared with the Gaussian and $\chi^2$ distributions. We also observe that the results of the Gaussian and the $\chi^2$ distributions are similar. However, the Gaussian distribution is found to be better overall since agreement to data is better at larger times. The close agreement of the Gaussian distribution approach gives confidence on the accuracy of the model predictions. Note that there are some measurements of the water table height (for example in Figs.~\ref{fig:matching}(c) and (d)) which are clear outliers. 
\rv{Regarding figures 3c and 3d, one can observe a “small cluster” of outliers at a certain water table level and distance from the dam (same at both cases), not appearing in figures 3a and 3b, which are probably due to various reasons: drought in that region, geology – impermeability of the bottom rock, changing characteristics of the aquifer soil in that region with time, which is not included in the model formulation.}

In the Germasogeia aquifer, the permeability of the porous medium varies spatially along the aquifer (solid red curve in Fig.~\ref{fig:permeability}(a)). In Fig.~\ref{fig:permeability}(a) we plot the water table height $\hat{h}$ and in Fig.~\ref{fig:permeability}(b) we plot $\hat{u}$, calculated using our model for $\hat{t}=0.01$ (one month after the first measurement), for constant permeability (spatially averaged value) and for the varying measured permeability.

We find that the discrepancy in the water table height and in the velocity is not significant (maximum deviation of 7.4\%), and hence the constant permeability assumption, in Fig.~\ref{fig:matching}, is well justified. We note that in addition to incorporating a spatially varying permeability, the model is valid also for spatially varying porosity, as well as for the temporal variation of these quantities, although this is not analysed in this work.

\clearpage
\newpage

\section{Water management and optimal recharge strategy  \label{opt_eq}}

The aquifer management issues we address here are sea water intrusion and water wastage. The types of aquifers we consider are managed with recharge points which pump water into the aquifer, thereby maintaining the water table. If the water table in the aquifer falls below a certain level, often due to excessive extraction or insufficient recharge, sea water can intrude,  making it unsuitable for human consumption. The minimum water table height was determined by WDD using measurements from the Germasogeia aquifer. 
On the other hand, excessive recharge of the aquifer is a waste of the clean water and resources.
In this study, using the recharge points (varying the recharge amount, location and number of points) to control the water table, we seek to minimise the amount of water needed to maintain the water table above the critical level. We use the optimisation procedure and data from the Germasogeia aquifer to investigate optimal recharge strategies but the methodology is valid for any sloping, sufficiently wide, long and thin aquifer.


\subsection{Formulating the optimisation problem}

There are a number of recharge points and we shall denote the recharge strength associated with each point by $R_i\un{m^3/day}$. The objective of our water management is to minimise the total recharge $\sum\limits_{i} R_i \un{(m^3/day)}$, balancing against the extraction and maintaining the minimum water level to prevent sea-water intrusion. 
We consider scenarios in which the strength of the recharge points is held constant for a long period of time such that the water table reaches its steady state. Hence, our optimal recharge strategy applies to the equilibrium water table, rather than an instantaneous water table. \rv{In other situations where the recharge points vary their strength much more rapidly, a time-dependent optimisation approach is more suitable}.

The recharge sources are considered as Gaussian distributions, as they have been shown to have the best agreement with data (see Section \ref{dupuit-forch}). 
In addition to the input recharge rates $0 \leq \hat{R}_i \leq \hat{R}_{max}$ (in non-dimensional terms), which are optimisation variables, we also allow the positions of the recharge points $0\leq\hat{x}_i\leq1$ to vary, as this may yield further water savings. In reality, this would require installing new recharge points, but the savings in water could potentially outweigh the cost. 
We also consider adding new recharge points. The water table $\hat{h}$ satisfies the governing equation (\ref{govh}) in the steady state (no time derivative term), and boundary conditions (\ref{nondimbc1}) and (\ref{nondimbc2}). Furthermore, for the purpose of this optimisation, we restrict our attention to the case where the permeability is constant, $\hat{k}=1$.
Finally, we impose a constraint on the water table height $\hat{h} \geq \hat{h}_{min}$, where $\hat{h}_{min}$ represents the critical (minimum) water level.

Therefore, to summarise, the mathematical description of our optimisation problem (in non-dimensional form) is
\beq
\min_{\{ \hat{R}_i, \,\hat{x}_i,\, \hat{h} \}} \sum\limits_{i=1:n_r} \hat{R}_i,\label{obj}
\eeq
subject to the constraints 
\begin{align}
\rvs{\frac{\partial}{\partial \hat{x}}\lb \hat{h} \lb 1-\frac{\partial\hat{h}}{\partial \hat{x}}\rb\rb=\phi \hat{s}\hat{h}}\quad & \rvs{\mathrm{for}\quad \hat{x}\in(0,1),} \label{con1}\\
\hat{h} \lb 1-\frac{\partial\hat{h}}{\partial \hat{x}}\rb=\hat{q} \quad &\mathrm{at}\quad x=0,\label{con2}\\
\hat{h}=\hat{H} \quad &\mathrm{at}\quad \hat{x}=1,\label{con3}\\
\hat{h}-\hat{h}_{min}\geq 0 \quad &\mathrm{for}\quad \hat{x}\in(0,1),\label{con4}\\
0\leq \hat{R}_i \leq \hat{R}_{max} \quad &\mathrm{for}\quad i=1,\ldots, n_r,\label{con5}\\
0\leq \hat{x}_i \leq 1 \quad &\mathrm{for}\quad i=1, \ldots, n_r,\label{con6}
\end{align}
where $n_r$ is the total number of recharge points.
We solve the optimisation problem numerically by discretising $\hat{h}$ into $N$ points. To solve equation (\ref{con1}) we use a second-order finite differences central scheme in space. For the boundary condition (\ref{con2}), we use a second-order forward scheme. 
We use the interior point method \citep{nocedal2006numerical} (with the IpOpt library \citep{wachter2006implementation}) to solve the non-linear constrained optimisation problem. Gradients are calculated using automatic differentiation in the JuMP package \citep{dunning2017jump} of the Julia programming language \citep{bezanson2017julia}. 

\subsection{Application to the Germasogeia aquifer with fixed recharge locations}

Having outlined the optimisation approach above, we now discuss how to use this to identify improved management protocols for the Germasogeia aquifer. For the Germasogeia aquifer, there are $n_r=4$ recharge points and the maximum recharge rates (due to tap capacity) are approximately $R_{max}=15000 \un{m^3/day}$ from each recharge point (giving a total of $60000\un{m^3/day}$ maximum recharge). 
Firstly, we take the locations of the recharge points as their field locations, as in Section 2. Later on, we also consider variable locations.
As discussed in Section 2, extraction is from 19 different locations along the aquifer. These locations lie between $x=1864\un{m}$ and $3889\un{m}$ (or $\hat{x}=0.33, \,0.68$). 
We consider each of them as point sinks, as described in Section 2. We choose extraction data from two periods which are characterised by high and low extraction rates, respectively. The first set of data is from November 2013 with a total extraction rate of $30035\un{m^3/day}$, and the second is from February 2015 with a total extraction rate of $14795\un{m^3/day}$. Using these two examples we develop an understanding of how to manage the aquifer when there is large or small water demand. 
Finally, the measured dam seepage rates are $2000-6000\un{m^3/day}$ (see \ref{appA}). We consider rates between $500\un{m^3/day}$ and $9000\un{m^3/day}$ to account for and study extreme situations (e.g. drought or flooding).

\begin{figure}[htbp!]
\centering
\begin{overpic}[width=0.45\textwidth]{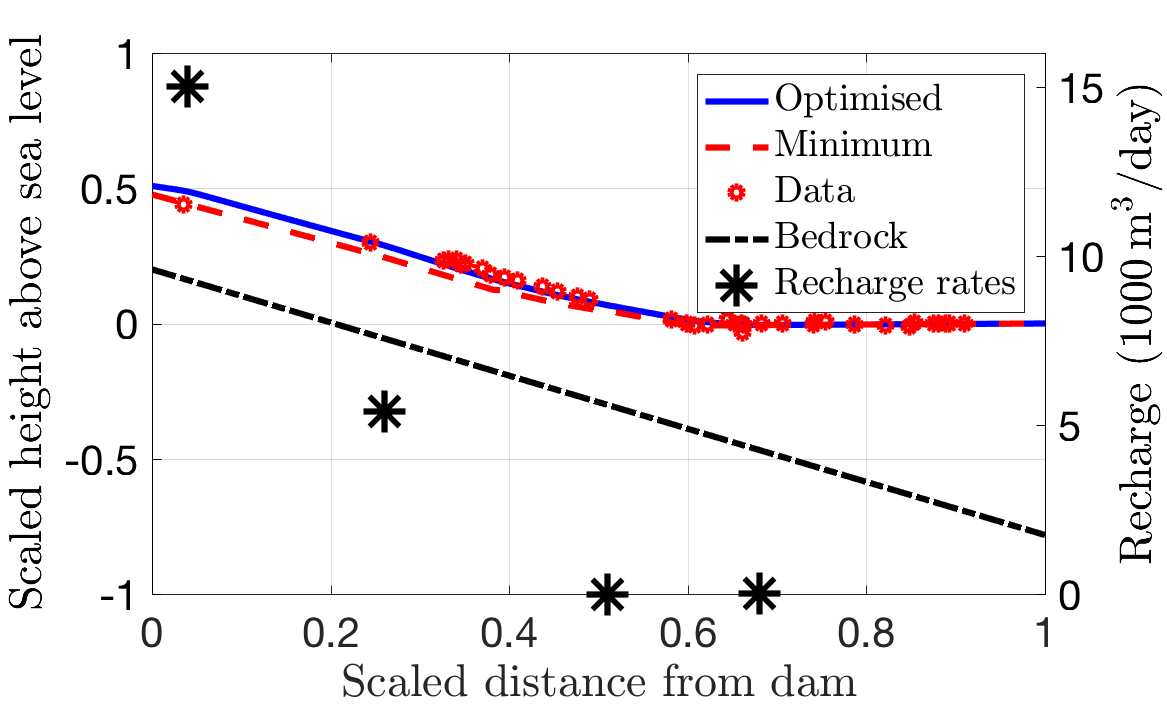}
\put(-5,60){(a)}
\end{overpic}
\hfill
\begin{overpic}[width=0.45\textwidth]{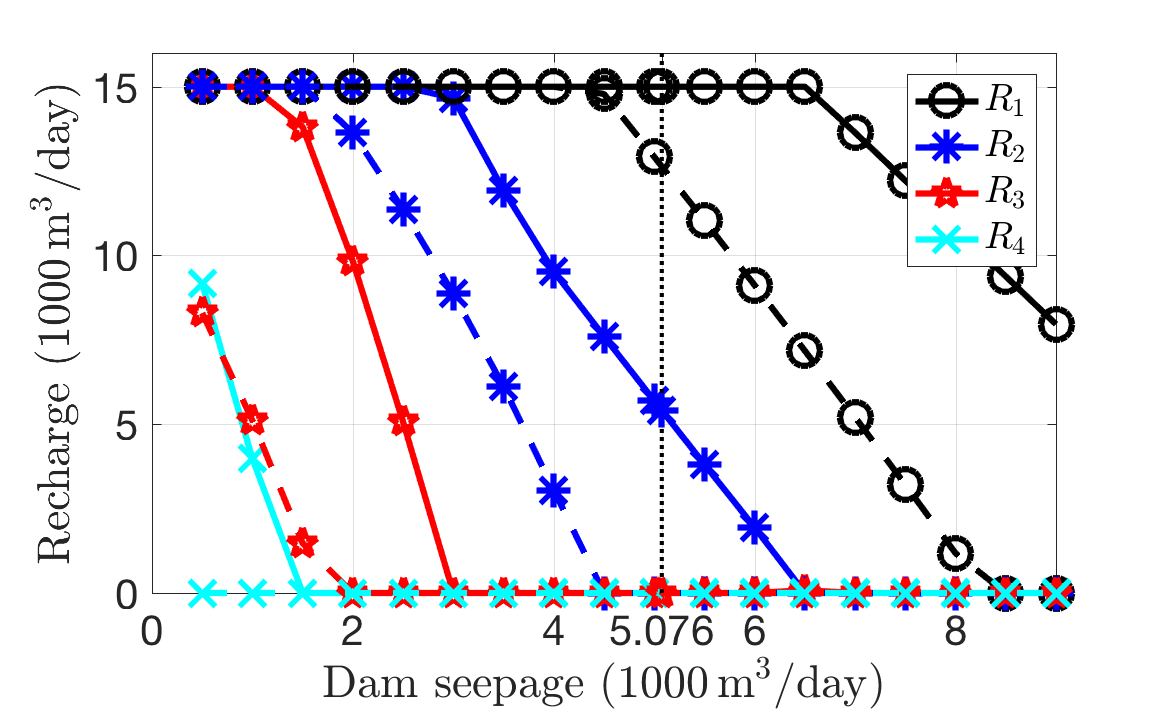}
\put(-5,60){(b)}
\end{overpic}
\caption{(a) Minimum recharge rates (right axis) applied to maintain the water table above the critical level. Dam seepage $5076\un{m^3/day}$ and extraction data $30035\un{m^3/day}$ taken from the Germasogeia aquifer in November 2013. Measured water table data and impermeable bedrock also shown. (b) Optimal recharge rates $R_i$ for various different dam seepage rates and extraction rates corresponding to both typical high value $30035\un{m^3/day}$ (solid) and typical low value $14795\un{m^3/day}$ (dashed). Data from November 2013 (a) shown as a vertical dotted line.
}
\label{fig1}
\end{figure}

Firstly, as an illustrative example, we optimise the recharge protocol for the dam seepage rate in November $2013$, which was $5076\un{m^3/day}$ (as interpreted from the dam water level data, refer to Fig. \ref{fig:timeseries_data}a). Using this value, together with the extraction data measured in that period, we solve equations (\ref{obj})$-$(\ref{con6}) to find the optimal recharge rates. In Fig. \ref{fig1}(a) we show the non-rotated optimised water table (solid blue line), for the optimised recharge rates, together with the minimum allowed water table (dashed red line), and the measured water table data for that period (red circles). We also plot the impermeable bedrock (dash-dotted black line), approximated with a linear slope. 
The optimal recharge rates are shown as black stars (refer to the values on the right hand axis). We see that the optimal recharge rates decrease with $x$, suggesting that pumping upstream is more important than downstream, and the final two points are not used at all. The total recharge amount in this example is $\sum \limits_{i=1:4} R_i = 20699 \un{m^3/day}$, which is $69\%$ of the total extraction rate. Therefore, if we choose a water management strategy which recharges exactly the same amount of water as is extracted, then we would be wasting $31\%$ of the equivalent total extraction rate. We can clearly see that the optimised water table stays above the critical level, as it is supposed to, and is slightly higher than the measured water table near the start of the aquifer. This is due to large recharge rates from $R_1$ and $R_2$ which are sufficient to maintain the water table further downstream above the critical level.

The next step is to see how variation in the dam seepage and extraction rates affect the optimal recharge strategy. In Fig. \ref{fig1}(b) we plot the optimal recharge rates for each individual recharge point, shown with four different marker styles, whilst varying both the dam seepage (horizontal axis) and extraction rates. High extraction rates (solid lines) are taken from November $2013$ data and low extraction rates (dashed lines) are taken from February $2015$ data. For both sets of extraction data, there is a common trend in that the optimal recharge rates $R_i$ decrease with increasing dam seepage. This is expected as, with more seepage from the dam, there is less need to recharge in order to maintain the water table above the critical level. For small extraction rates, each recharge point has a lower optimal value than for the larger extraction rate. This is expected because, if more water is being extracted from the dam, we must account for the loss of water with extra recharge. Another observation is that there is a hierarchy of importance for the recharge points, with $R_1$ (first) having the largest flow rate and $R_4$ (last) having the smallest. Therefore, above some critical dam seepage values, \rv{recharge points $R_2$, $R_3$, and $R_4$} can be turned off completely. A line is drawn in the figure to indicate a dam seepage rate of $5076 \un{m^3/day}$, which corresponds to Fig. \ref{fig1}(a). 

\begin{figure}
\centering
\begin{subfigure}{0.45\textwidth}
\centering
\begin{overpic}[width=1\textwidth]{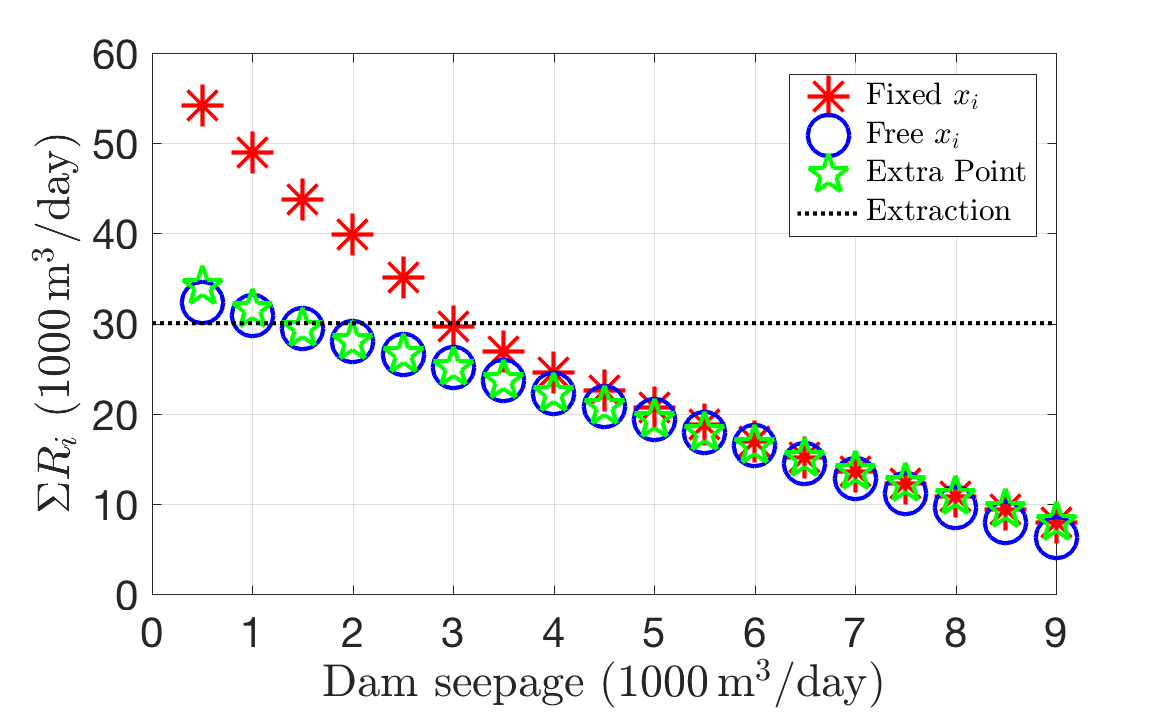}
\put(-5,55){(a)}
\end{overpic}
\end{subfigure}
\begin{subfigure}{0.45\textwidth}
\centering
\begin{overpic}[width=1\textwidth]{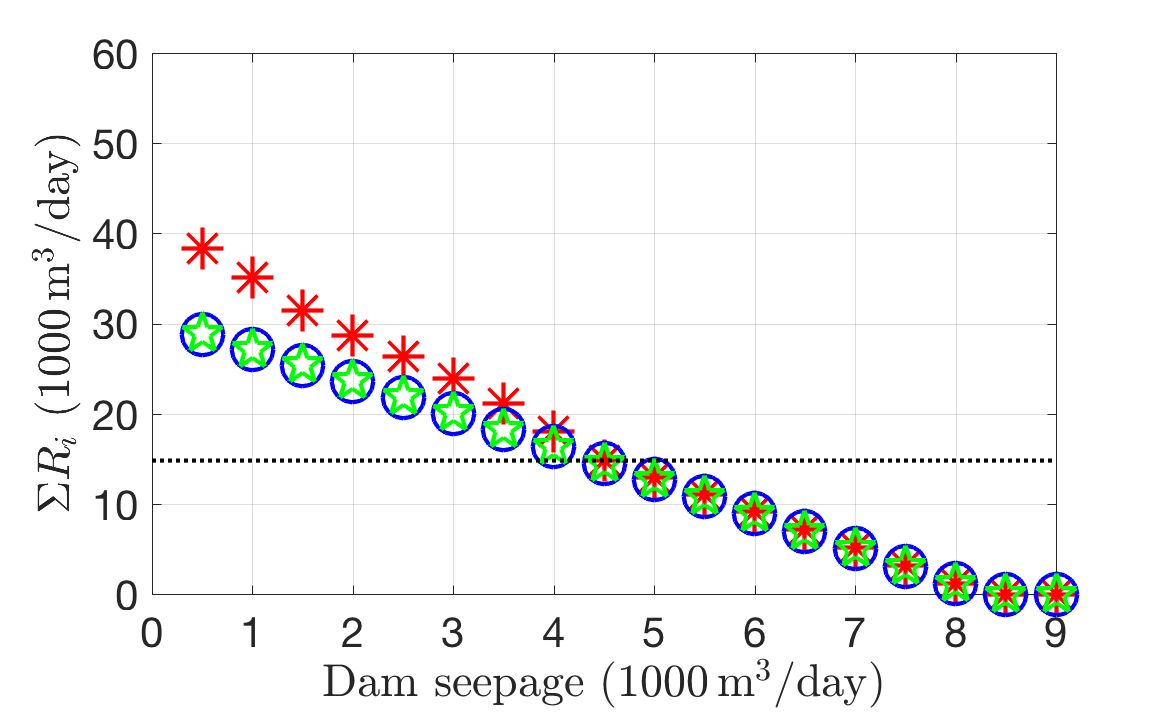}
\put(-5,55){(b)}
\end{overpic}
\end{subfigure}
\caption{Comparison of optimised total recharge (\ref{obj}) for the case where the locations of the $4$ recharge points are fixed, the case where we include their locations as optimisation variables, and the case where we take the locations as fixed but add a fifth recharge point which has free location. (a) High extraction $30035\un{m^3/day}$. (b) Low extraction $14795\un{m^3/day}$.
}
\label{fig2}
\end{figure}

\subsection{Application to the Germasogeia aquifer with variable recharge locations}

If we now allow the location of the recharge points, $x_i$, to vary, it is interesting to see whether we can achieve more water savings. In Fig. \ref{fig2} we plot the total recharge rate for when $x_i$ are fixed at their real values (red line with asterisks), and when $x_i$ are free variables (blue line with circles). We also consider the four recharge locations fixed and introduce an extra recharge point $R_5$ as a free variable (green line with asterisks). For both (a) high extraction rates and for (b) low extraction rates we explore the effect of variable recharge location(s). Both cases, either having low dam seepage rates and allowing variable recharge point locations, or adding a fifth point, result in a significant contribution to water savings. 
Water quantities of up to $9478 \un{m^3/day}$ and $21860 \un{m^3/day}$ can be saved for the low extraction regime and high extraction rates, respectively. Interestingly, the flexibility of an additional recharge point has almost the same contribution to water savings as allowing the remaining other recharge points to change location. For high seepage rates, the effect of variable recharge point locations or a new recharge point on the water saving is minimal. This is because the only active recharge point is $R_1$, which is located very close to the dam at $x=0$. When we allow the locations to vary freely, the locations tend to cluster towards small values of $x$ because, as we have seen, these are the most important. Thus, at large dam seepage rates there is a little advantage in relocating the existing recharge points, or installing a new one.  However, for the Germasogeia aquifer, dam seepage rates are low and an optimised recharge protocol becomes more important.

In Fig. \ref{fig2}, we also indicate the total extraction rate (horizontal line), corresponding to both the low and high extraction cases. For high extraction rates ($30035 \un{m^3/day}$), if we do not constrain the recharge point locations, or add a new recharge point, then a total recharge rate equal to the total extraction rate is sufficient to maintain the water table above its critical value (except for extremely small dam seepage $\sim500\un{m^3/day}$). However, when the recharge points are fixed at their current locations, dam seepage rates below $\sim3000 \un{m^3/day}$ require recharge rates much larger than the total extraction rate. For the low extraction case ($14795 \un{m^3/day}$), it is clear that for dam seepage rates smaller than $\sim4500 \un{m^3/day}$, the total recharge must be larger than the total extraction, irrespective of the recharge point locations. Therefore, though it may seem reasonable that a total recharge rate equivalent to the total extraction rate should be sufficient to maintain the critical water level, we have seen here that this is not always the case. In particular, for small dam seepage rates it is often necessary to recharge as much as double the total extraction rate (see the left hand points in Fig. \ref{fig2}(b)). However, by relocating the recharge points, or adding a new point, it is possible to greatly reduce the necessary recharge rates in these situations, thereby saving water.

Although we have restricted our attention here to the optimisation of the equilibrium water table, it would be interesting to investigate how our findings would be different in the case of a time-dependent optimisation of the dynamic water table. However, we leave this to the list of future work.

\section{Modelling sea water intrusion} \label{sec:sea_intru}

In the case of drought or excessive extraction from the aquifer, we expect the water table to drop, thereby drawing heavier salty water from the sea into the aquifer. 
The interface between the fresh water and the sea water is thin because the relevant Peclet number is large, and the transport of salt is advection-dominated. The Peclet number Pe$_s=KH/D_s$ is given in terms of the hindered diffusivity $D_s=\phi D_{s_0}$, where $D_{s_0}$ is the bulk diffusivity (which we assume isotropic). Therefore, if we take the bulk diffusivity of salt in water as $D_{s_0}\approx 10^{-9}\,\mathrm{m^2/s}$ \cite{cussler2009diffusion} 
and, considering the characteristics of the Germasogeia aquifer $\phi=0.4$, $K=1.5\times 10^{-3}\,\mathrm{m/s}=130~\un{m/day}$ (spatially averaged) and $H=80\,\mathrm{m}$, then Pe$_s\approx 3\times10^8$. Since, the transition region between sea water and fresh water is thin, we model it as a sharp interface between two fluids (with different densities). This reduces the computation load significantly as we don't have to solve an additional transport equation. 
We introduce the non-dimensional density function
\begin{equation}
\hat{\rho}=\begin{cases}
1 &\quad \mathrm{in}\,\mathrm{fresh}\,\mathrm{water},\\
1+\beta  &\quad \mathrm{in}\,\mathrm{sea}\,\mathrm{water},  
\end{cases}
\end{equation}
and the viscosity function
\begin{equation}
\hat{\mu}=\begin{cases}
1 &\quad \mathrm{in}\,\mathrm{fresh}\,\mathrm{water},\\
1+\gamma  &\quad \mathrm{in}\,\mathrm{sea}\,\mathrm{water},  
\end{cases}
\end{equation}

where $\beta$ and $\gamma$ are non-dimensional constants relating the densities and viscosities of sea water (and which can be related to salinity).
We model the flow in the aquifer using the two-dimensional, unsteady Darcy-Brinkman equations \cite{furman2008modeling, joodi2010development, chen2010asymptotic, neale1974practical}, which include viscous effects. Unlike the Dupuit-Forchheimer model in Section \ref{dupuit-forch}, this allows us to impose appropriate stress conditions on the sharp interface. 
As before, for the sake of simplicity, we consider constant permeability $\hat{k}=1$. In non-dimensional form, the continuity and the Darcy-Brinkman equations are

\begin{align} 
\phi \hat{\rho} \lb \frac{\partial \hat{u}}{\partial \hat{x}}+ \frac{\partial \hat{w}}{\partial \hat{z}}\rb  &= \hat{s}, \label{darcybrink1}\\
\frac{{\rm Fr}^2}{\epsilon} \hat{\rho} \frac{\partial \hat{u}}{\partial \hat{t}} + \hat{\mu}\phi \hat{u} &= -\frac{\partial \hat{p}}{\partial \hat{x}} + \hat{\rho} +\hat{\mu}\frac{{\rm Fr}^2}{\epsilon^2 {\rm Re} } \lb \frac{\partial^2 \hat{u}}{\partial \hat{x}^2} + \frac{1}{\epsilon^2} \frac{\partial^2 \hat{u}}{\partial \hat{z}^2} \rb , \label{darcybrink2}\\	
\frac{{\rm Fr}^2}{\epsilon} \hat{\rho} \frac{\partial \hat{w}}{\partial \hat{t}} + \hat{\mu}\phi \hat{w} &=-\frac{1}{\epsilon^2}  \frac{\partial \hat{p}}{\partial \hat{z}} -  \frac{\hat{\rho}}{\epsilon^2} +\hat{\mu}\frac{{\rm Fr}^2}{\epsilon^2{\rm Re}} \lb \frac{\partial^2 \hat{w}}{\partial \hat{x}^2} + \frac{1}{\epsilon^2} \frac{\partial^2 \hat{w}}{\partial \hat{z}^2} \rb ,\label{darcybrink3}
\end{align}

where Re=$\rho K H / \mu$ is the Reynolds number and Fr$^2=K^2 H^2 / (g L^3)$ is the Froude number squared. Unlike the Dupuit-Forchheimer approximation in Section \ref{dupuit-forch}, here we do not neglect terms of order $\mathcal{O}(\epsilon)$ or higher.

We impose the boundary conditions (\ref{nondimbc1})$-$(\ref{nondimbc2}) and additional conditions at the sharp interface between sea water and fresh water, which is denoted by $\hat{z}=\hat{h}_i(\hat{x},\hat{t})$, as follows:

\begin{flalign} \label{eq:fresh_salt_water_interface}
\text{continuity~of~mass:} \hspace{4cm} ~\hat{\rho}_1 \mathbf{\hat{u}}_1 &= \hat{\rho}_2 \mathbf{\hat{u}}_2,
\\
\text{continuity~of~stress:} \hspace{2cm} ~\hat{p}_1 \mathbf{I} + \frac{{\rm Fr}^2}{\epsilon^2 {\rm Re}} ~ \nabla \mathbf{\hat{u}}_1   &= \hat{p}_2 \mathbf{I} + \frac{{\rm Fr}^2}{\epsilon^2 {\rm Re}} \frac{\mu_2}{\mu_1} ~ \nabla \mathbf{\hat{u}}_2,
\end{flalign}

where $\mathbf{u} = (u,w)$. The index 1,2 represents the sea and fresh water characteristics and $\mathbf{I}$ represents the identity matrix. On the aquifer base $\hat{z}=0$, in addition to the impermeability boundary condition (\ref{imp}), we should also impose the no-slip condition
\beq
\hat{u}=0\quad \mathrm{on}\quad \hat{z}=0.
\eeq

As initial condition for the interface $\hat{h}_i$, we consider a vertical interface at the sea front ($\hat{x}=1$).

In addition, we impose the GH relationship, which relates the water table height above sea level to the depth of the sharp interface below sea level. If we denote the sea level height (in non-dimensional rotated coordinates) as $\hat{h}_s$, then the GH relationship is
\begin{equation} \label{eq:GH_relation}
\lb 1+\beta \rb \lb \hat{h}_s - \hat{h}_i \rb = \hat{h} -\hat{h}_i.
\end{equation}

We solve the problem \eqref{darcybrink1}--\eqref{eq:GH_relation}
to analyse the effect of sea water intrusion, choosing two cases to investigate. We discuss the numerical techniques used to solve this system in \ref{appB}.
The first scenario corresponds to severe sea water intrusion. This is achieved by setting an initially very low water table, below the sea water level. Furthermore, we consider no recharge or extraction from the aquifer, and the dam seepage is set to minimal, $Q=100~\un{m^3/day}$ . We have chosen this scenario in order to illustrate the isolated effect of having an excessively low water table. 

In the second scenario, we use three different values for the dam seepage rate, $Q=500$, $2500$ and $12500\,\mathrm{m^3/day}$, respectively for a drought period, a moderate rainfall period, and a flooding period. We use constant extraction rates, given by the Germasogeia aquifer management data for October 2013, whereas we keep the recharge sources switched off to illustrate the effect of excessive extraction. In this case, the initial water table height is set as a polynomial fit with the observation data from October 2013 (as discussed in Section 2).  
In both scenarios, we assume constant permeability in the aquifer.
Additionally, the sharp interface between sea water and fresh water is initially vertical, set at $\hat{x}=1$. 

\begin{figure}
\centering
\footnotesize
\begin{overpic}[width=1\textwidth]{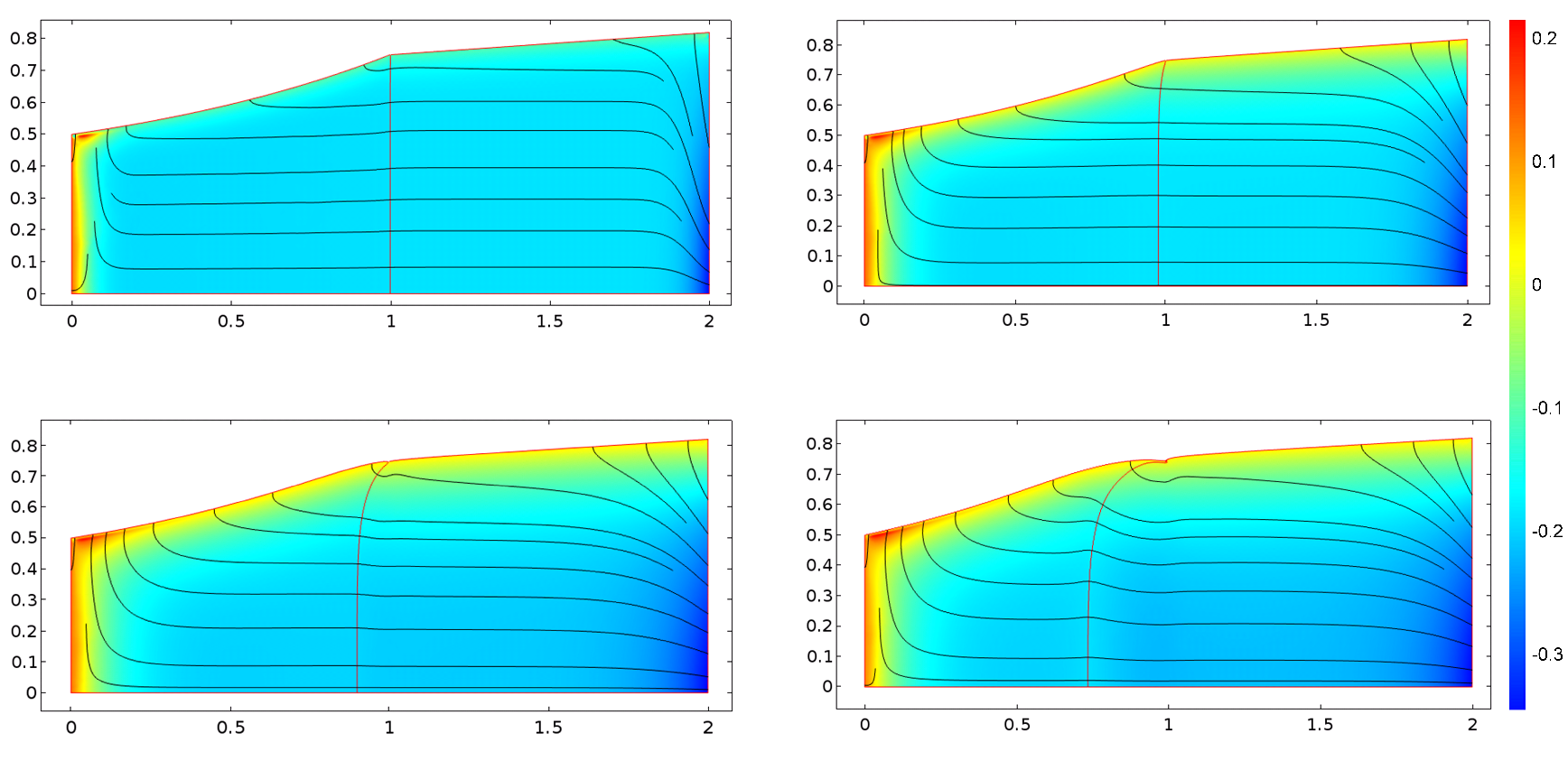}
\put(4,45){(a)}
\put(55,45){(b)}
\put(4,19){(c)}
\put(55,19){(d)}
\put(24,0){$\hat{x}$}
\put(74,0){$\hat{x}$}
\put(24,26){$\hat{x}$}
\put(74,26){$\hat{x}$}
\put(-2,13){\rotatebox{90}{$\hat{z}$}}
\put(-2,39){\rotatebox{90}{$\hat{z}$}}
\put(49,13){\rotatebox{90}{$\hat{z}$}}
\put(49,39){\rotatebox{90}{$\hat{z}$}}
\put(100,25){$\hat{u}$}
\put(15,35){\textbf{Fresh}}
\put(30,35){\textbf{Sea}}
\put(15,10){\textbf{Fresh}}
\put(30,10){\textbf{Sea}}
\put(65,35){\textbf{Fresh}}
\put(80,35){\textbf{Sea}}
\put(60,10){\textbf{Fresh}}
\put(80,10){\textbf{Sea}}
\end{overpic}
\caption{[to be viewed in colour] Sea water intrusion for an initial water table set well below sea level, with no extraction or recharge, and dam seepage set to $100~\un{m^3/day}$. Different times are displayed (a) $\hat{t} = 0$, (b) $\hat{t} = 0.2$ (1.6 years), (c) $\hat{t} = 0.37$ (3 years) and (d) $\hat{t} = 0.87$ ($6.9$ years) 
The background colour represents the magnitude of the horizontal velocity $\hat{u}$ and the black solid lines represent the flow streamlines. 
The vertical solid black curve at $\hat{x}=1$ in (a) denotes the sharp interface between fresh water and sea water, which intrudes inland slowly with time. }
\label{fig:sea_water_intru}
\end{figure}
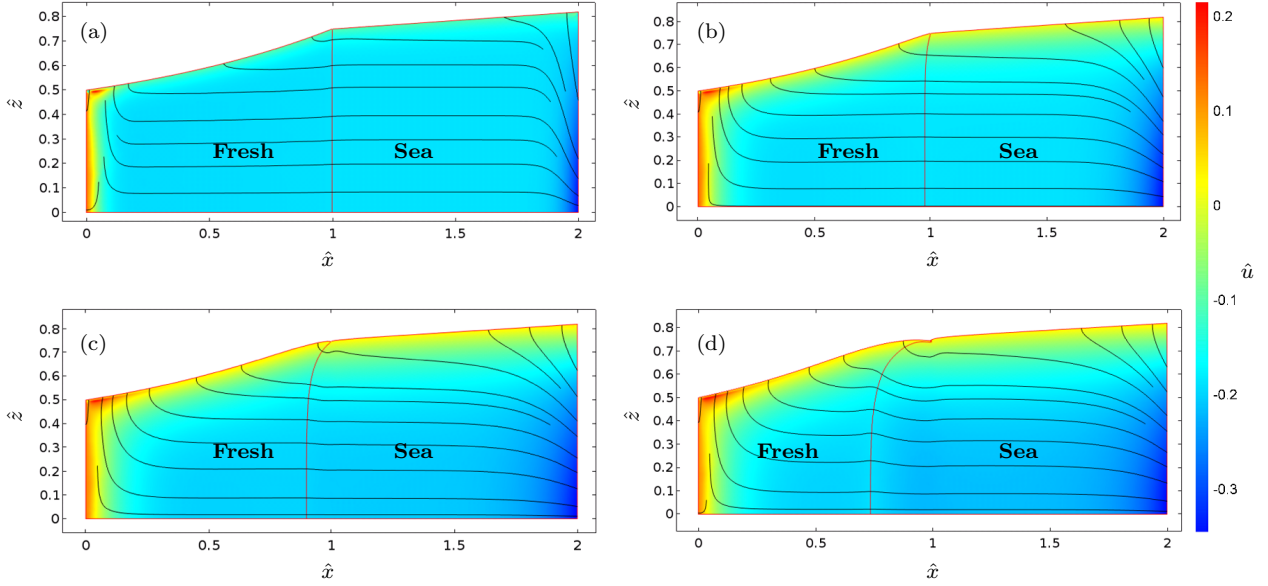

We plot the results of the first scenario in Fig.~\ref{fig:sea_water_intru} at different times $t=0$, $1.6$, $3$ and $6.9$ years after the initial condition. The background colour represents the magnitude of the fluid velocity in the horizontal direction $\hat{u}$. 
We can see that there are significant regions of negative velocity. In particular, the velocity is negative at the bottom of the sharp interface, indicating that sea water is being drawn into the fresh water region. 
At the top of the water table the velocity is positive, indicating that the fresh water is still draining into the sea, despite the backward flow beneath.
It can be observed that both the water table $\hat{h}$ in the aquifer region ($\hat{x} < 1$), and the sharp interface $\hat{h}_i$ change position significantly over $6.9$ years. 

Since sea water is denser than fresh water (with a density ratio of $1.03$), the base of the interface progresses faster than the top, as is evident from Figs.~\ref{fig:sea_water_intru}(c) and (d). The difference of the interface location from sea level to the aquifer base is more than 1 km after $6.9$ years.  

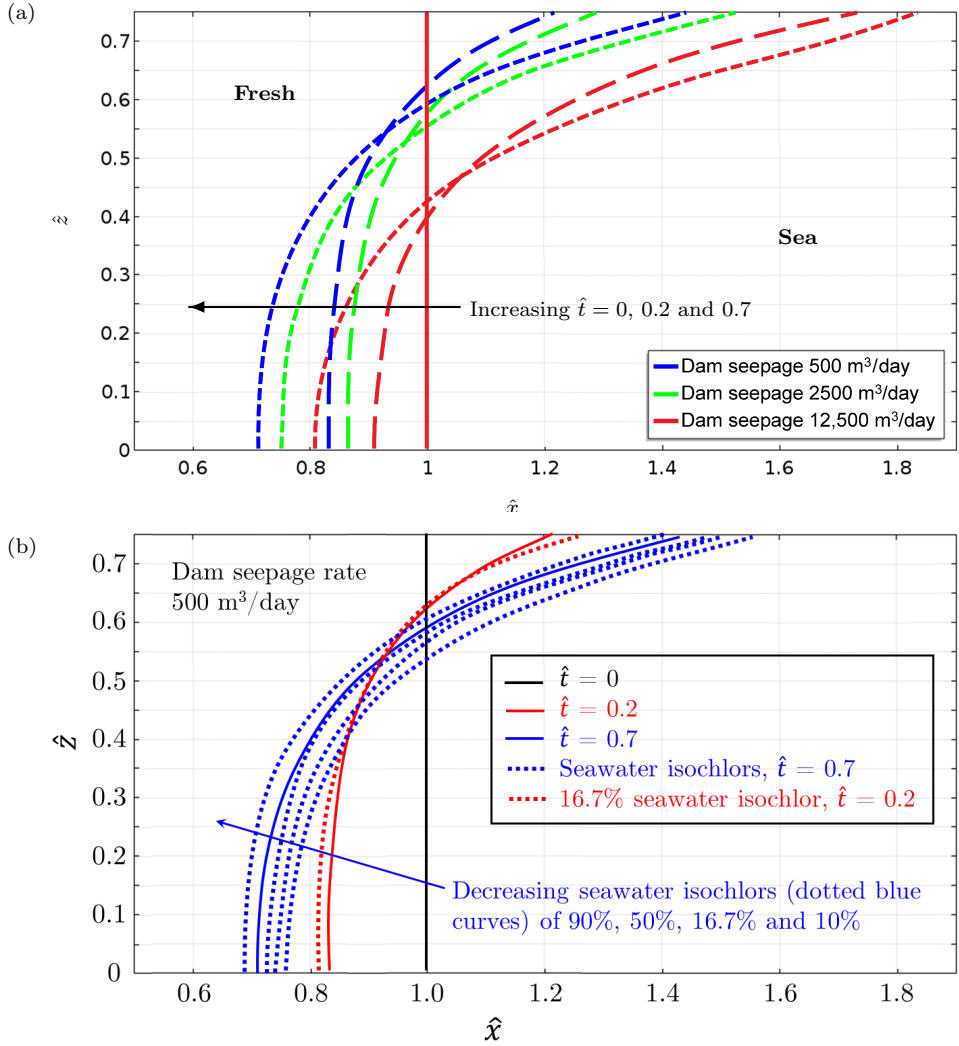
\begin{figure}
\centering
\footnotesize
\thicklines
\begin{overpic}[width=0.75\textwidth]{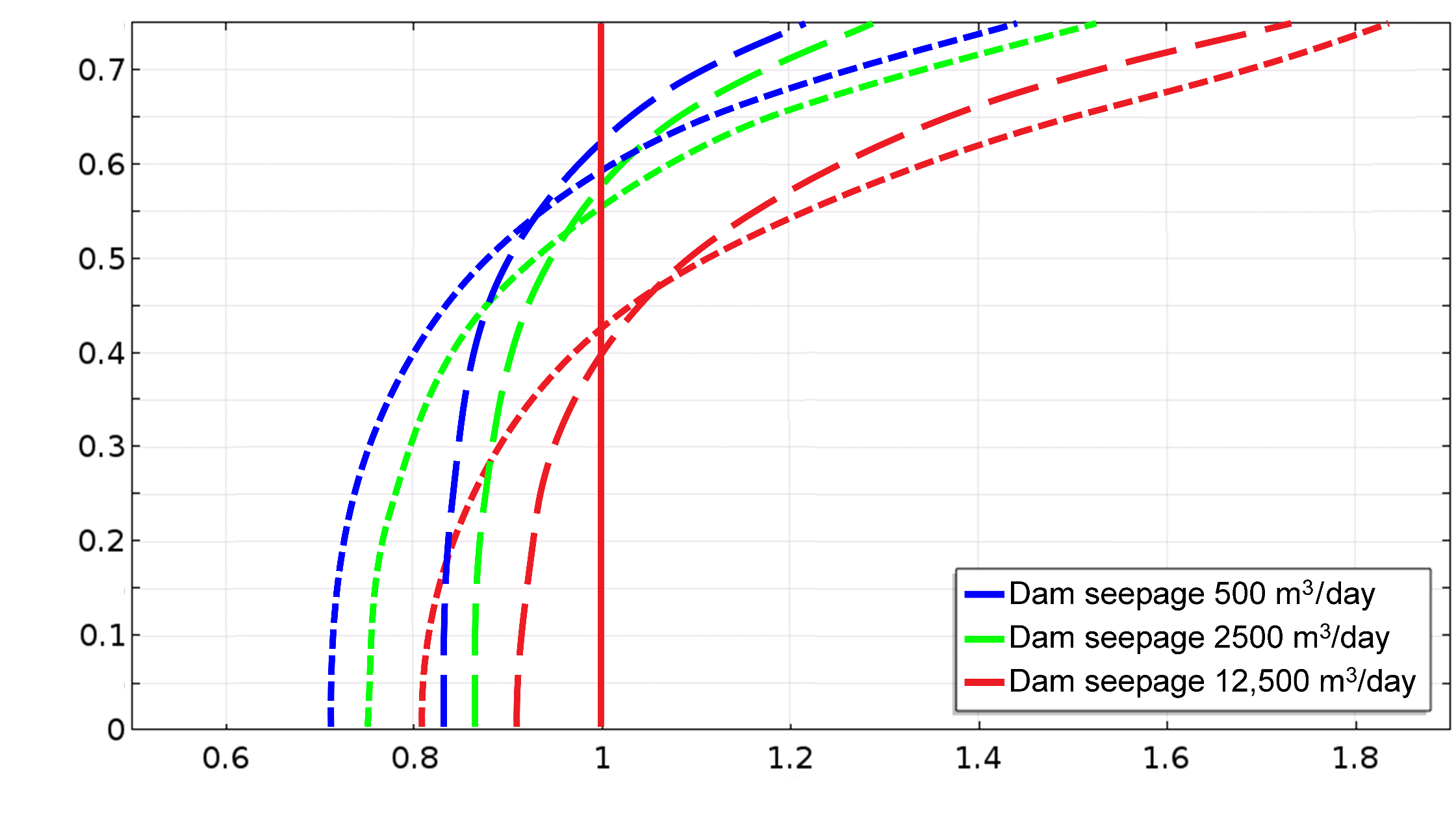}
\put(0,32){\rotatebox{90}{$\hat{z}$}}
\put(50,0){$\hat{x}$}
\put(20,46){\textbf{Fresh}}
\put(80,30){\textbf{Sea}}
\put(45,23){\vector(-2,0){30}}
\put(46,22){\footnotesize Increasing $\hat{t}=0$, 0.2 and 0.7}
\put(-5,55){(a)}
\end{overpic}

\begin{overpic}[width=0.75\textwidth]{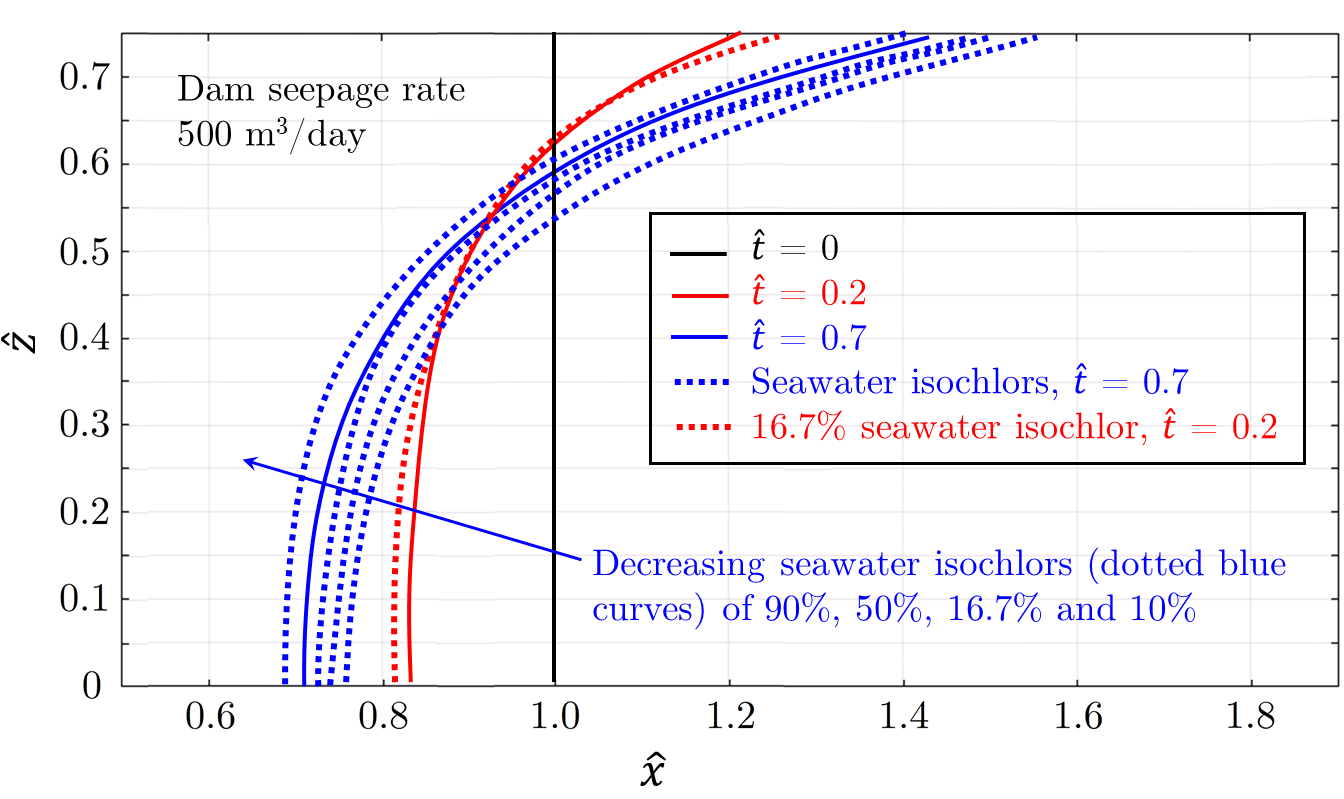}
\put(-5,55){(b)}
\end{overpic}
\caption{(a) Dynamics of the sharp interface between fresh water and sea water for different dam seepage rates. The dashed and the dotted lines correspond to $\hat{t}$ = 0.2 and 0.7.
(b) Comparison of the fresh-sea water interfaces with the open source SUTRA code and the current model results for the case of 500 $m^3$/day dam seepage rate. The dotted lines represent the result obtained using the SUTRA code while the solid line is from the proposed model calculations developed here. \rvs{The curves from the SUTRA code which are closest match to the present approach correspond to $\beta=0.005$ (16.7\% seawater isochlor). Additionally, some more curves are presented for $\hat{t}=0.7$, representing 90\%, 50\% and 10\% seawater isochlors, equivalent to $\beta=0.027$, 0.015 and 0.003.}
The extraction rates are constant and taken as their value in October 2013 ($\hat{t}=0$), whereas recharge rates are switched off. The water table is initially at the October 2013 measured location (polynomial fit) and the sharp interface is initially vertical, set at $\hat{x}=1$. 
}
\label{fig:sea_water_intrusion_2}
\end{figure}

Whilst the first scenario illustrates the effect of an excessively low water table, the second scenario focuses on the role of dam seepage in sea water intrusion. In Fig.~\ref{fig:sea_water_intrusion_2}a we display the position of the sharp interface between fresh water and sea water for the different dam seepage rates at two different time intervals, corresponding to 1.6 ($\hat{t}=0.2$) and 5.6 years ($\hat{t}=0.7$). 
In each case the sharp interface is initially at the same location and as time progresses the interface intrudes into the aquifer.
As expected, with smaller dam seepage rates the intrusion is much larger. For a dam seepage rate of $500 \un{m^3/day}$, after around 5.6 years, sea water encroaches almost $30\%$ of the aquifer (at the base).

\rv{
In order to have credibility and validate the results of the developed model, we have set up a similar model using the widely accepted saturated-unsaturated transport (SUTRA) variable-density ground-water flow package \mbox{\cite{voss2002sutra}}. The comparison of the results using these two approaches are presented in Fig.~\ref{fig:sea_water_intrusion_2}b for a specific value of the dam seepage rate of 500 $m^3$/day. It can be seen here that the results from the SUTRA code are close to the developed model calculations (within reasonable accuracy), \rvs{corresponding to $\beta=0.005$ which is equivalent to 16.7\% seawater isochlor}. The details of the model set up in the SUTRA package are described in \ref{appB}.
}


\clearpage
\newpage
\section{Conclusions and future work}
In this work we have developed a holistic framework for the management of sloping, long and thin, coastal aquifers. We have taken as a test case the Germasogeia aquifer in Limassol, Cyprus, and used data provided over a three-year period by the Cyprus Water Development Department.

We first developed a simple two-dimensional mathematical model, based on Darcy flow for porous media, for the water table height and the water velocity, neglecting sea water intrusion (Dupuit-Forchheimer approximation). The model was validated with the data provided by WDD. Spatial permeability variation and its effect on the water table has been studied and it has been shown that small variations do not have a strong effect on the water table. Since this is the case for the Germasogeia aquifer, we have adopted in all parts of this study a constant permeability. However, in future work the model can easily be extended to study spatially varying and time-dependent permeability as may be needed in other aquifer studies.

Using the steady version of our model, we subsequently developed an optimised recharge strategy and identified the optimal recharge rates for a desired extracted water volume while the water table height was maintained at an acceptable level. We considered various practical scenarios where we varied the rates and locations of the recharge sites (while keeping the extraction sites fixed) and showed that considerable water savings can be achieved as compared to the current empirical strategy followed by WDD. In future work, it would be of further interest to develop a time-dependent optimisation protocol for a dynamically varying water table, and to additionally explore the effect of variation of the extraction locations and rates.

We then considered the case of an accidental leakage and predicted the transport of pollutants in the aquifer for various scenarios of practical interest. We used an advection-diffusion equation for the pollutant and assumed that its velocity is given by the water velocity, as this was determined in Section 2. Three scenarios were considered for the duration of the leakage: non-stop leakage, and ``pulse" leakage lasting one day or 30 days. In the case of non-stop, unhindered and undetected leakage (worst case scenario) we found that the aquifer would be completely polluted in approximately three years. In the two cases of pulse leakage, the impact is less severe. Exploring also variation of the recharge rates we found that contamination can be flushed out faster as the recharge rates increase. As the Peclet number is very large (diffusion effects are very small) the contaminant propagation front is almost vertical and this fact suggests that a monitoring device would be expected to detect sudden changes in contaminant concentration levels.

Finally, we developed a time-dependent, two-dimensional model of saturated-unsaturated groundwater flow based on the Darcy-Brinkman equations, and we showed that the problem can be case as a two-fluids problem with a sharp interface since diffusion effects are very small. This framework allowed us to easily study sea water intrusion by tracking the position of the water table and of the seawater-freshwater interface. We applied the model in four scenarios of interest; initially a very low water table with no extraction or recharge and then for three potential cases-drought, moderate rainfall and flooding. We found that even in the case of drought (dam seepage rate of $500 \un{m^3/day}$) we expect sea water to encroach almost $30\%$ of the aquifer (at the base) in less than 6 years. The analysis done here offers an alternative, less computationally expensive methodology to solving the fully coupled density-concentration diffusion problem. 
\rv{This part of the work is validated using the popular SUTRA code for the variable density groundwater flow and the comparison of the results is also shown here.}
\rvs{Comparison of the computational resource requirement for this new approach is a subject of further investigation, which is outside the scope of the present study. However, the current work is aimed towards providing a foundation in the development of new alternative efficient technique.}

In the present study the width of the aquifer has been assumed infinite (no edge effects) and hence we could treat the problem as two-dimensional. As we have seen this is a good approximation for the Germasogeia aquifer for a significant part of the aquifer but we expect a closer agreement with data if we take the width variation into account. It is also expected that also in other aquifers we would not be able to neglect the effects of width variation.

\section*{Acknowledgements}
The authors would like to acknowledge the 125\textsuperscript{th} European Study Group with Industry (ESGI125, www.esgi-cy.org) which was held in Limassol, Cyprus, 5-9 December 2016. Dr K. Kaouri (Chair of the ESGI125 Organising COmmittee) would especially like to thank the  Mathematics for Industry Network (MI-NET, COST Action TD1409), KPMG and the Cyprus University of Technology for supporting this workshop. The authors are also thankful to the Cyprus Water Development Department (WDD), Limassol District Office, for proposing the problem, fruitfully collaborating with the Study Group participants and for financial support. The authors also acknowledge support from WDD that aided them to continue the work after the Study Group. The authors are also thankful to all the members of the ESGI125 team for stimulating discussions: Prof D. Papageorgiou (Imperial College London, UK), Dr Hilary Ockendon (University of Oxford, UK), Dr C. Nikolopoulos (University of the Aegean, Greece), Dr T. Ivanov (Sofia University, Bulgaria), Prof. P. Frolkovic (Slovak University of Technology, Slovakia), Mr A.N. Riseth (University of Oxford, UK), Prof. A. Lacey (Heriot-Watt University, UK) and Dr V. Rottsch{\"a}fer (Leiden University, Netherlands). We would also like to thank Prof G. Sander (University of Loughborough/University of Oxford) for helpful discussions.

G. Benham acknowledges the partial financial support from WDD and the University of Oxford in the second phase of the work, after the Study Group. Dr R. Mondal acknowledges financial support from MI-NET through a Short Term Scientific Mission (STSM) grant. Dr S. Mondal also acknowledges the partial support from WDD and the Royal Society towards travel and subsistence.

\vspace{2cm}

\appendix
\section{Time-series data of the recharge, extraction and dam seepage \label{appA}}

\begin{figure}[htbp!]
\centering
\footnotesize
\begin{overpic}[width=0.32\textwidth]{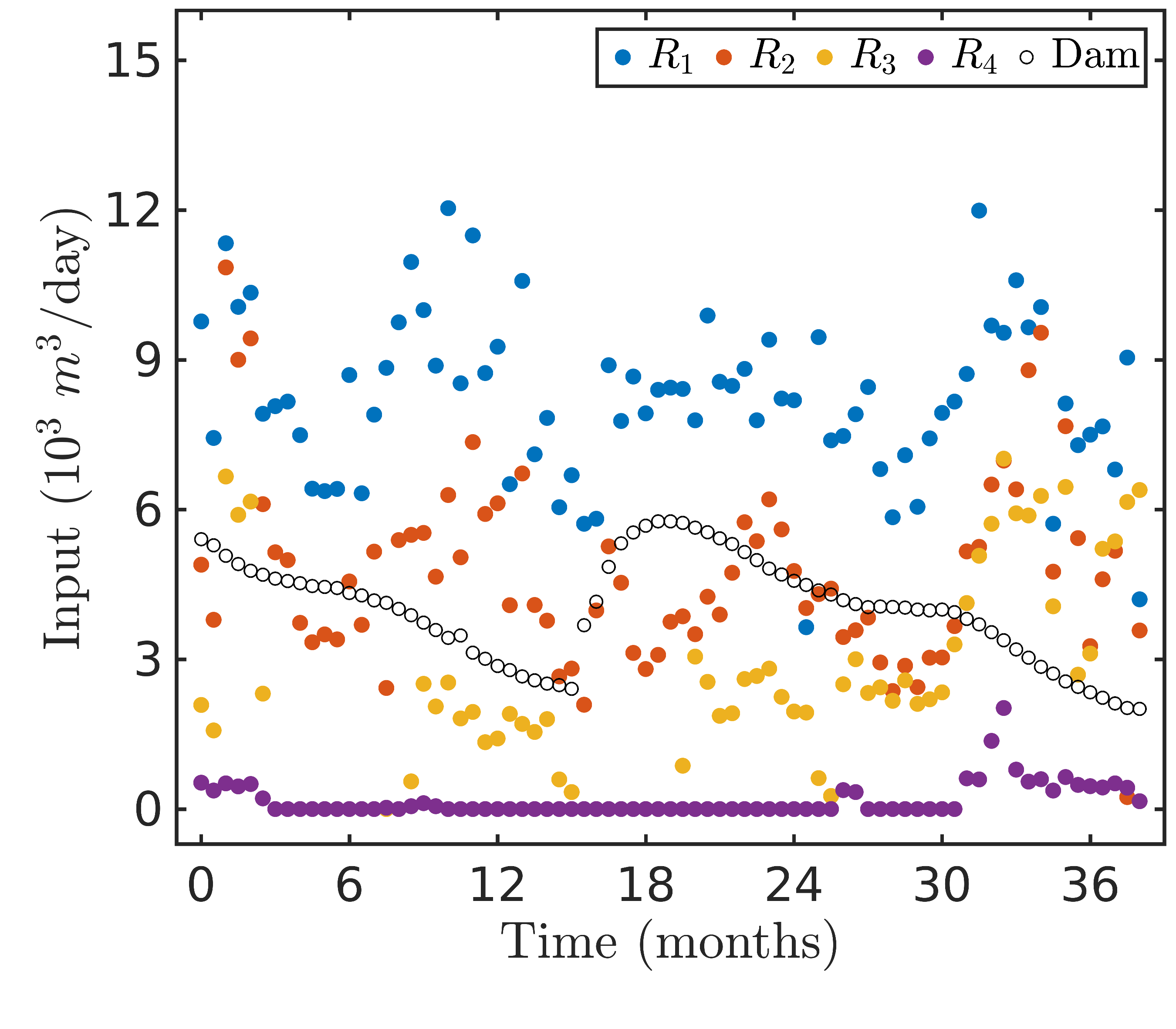}
\put(20,77){(a)}
\end{overpic}
\hfill
\begin{overpic}[width=0.32\textwidth]{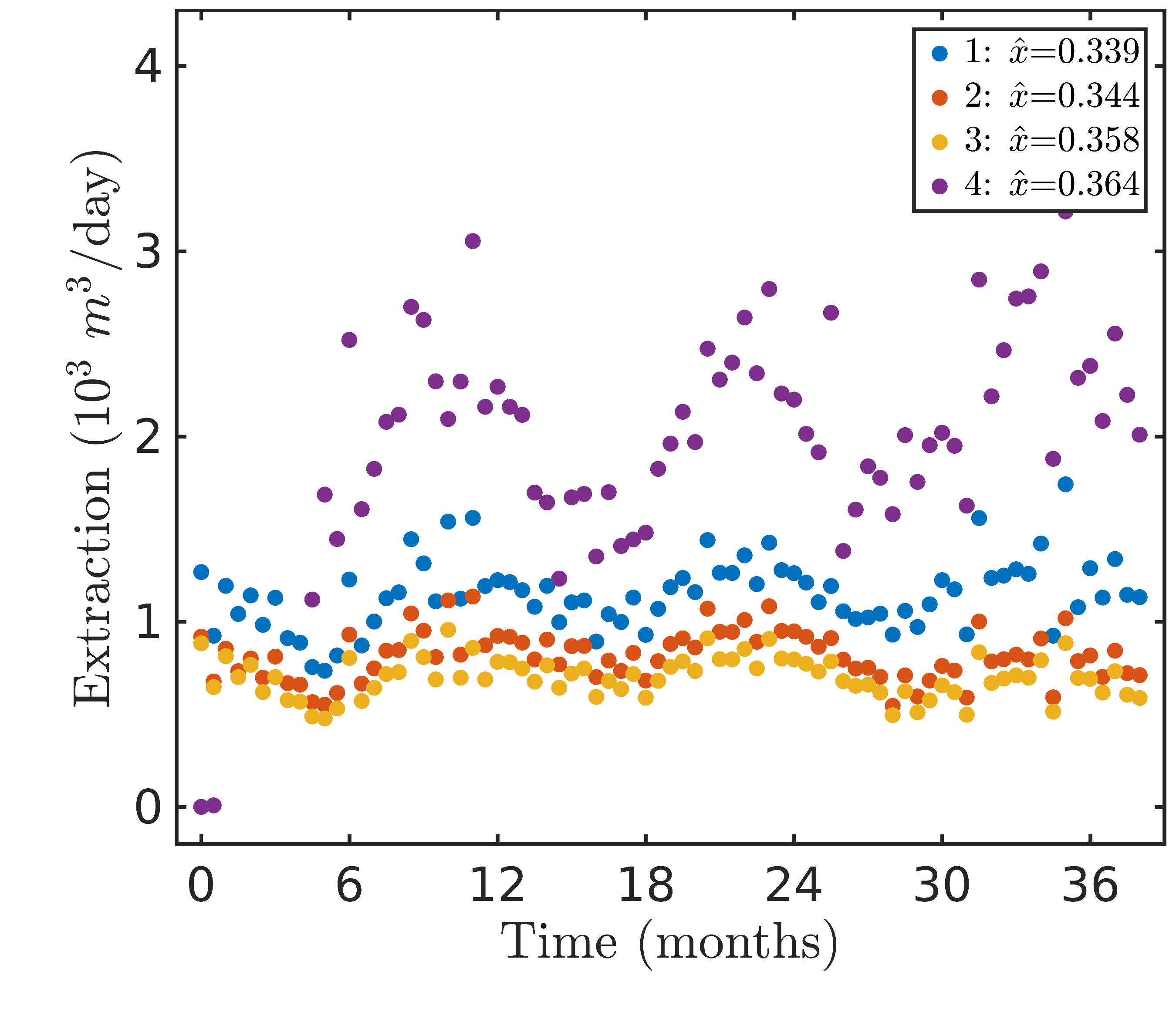}
\put(20,77){(b)}
\end{overpic}
\hfill
\begin{overpic}[width=0.32\textwidth]{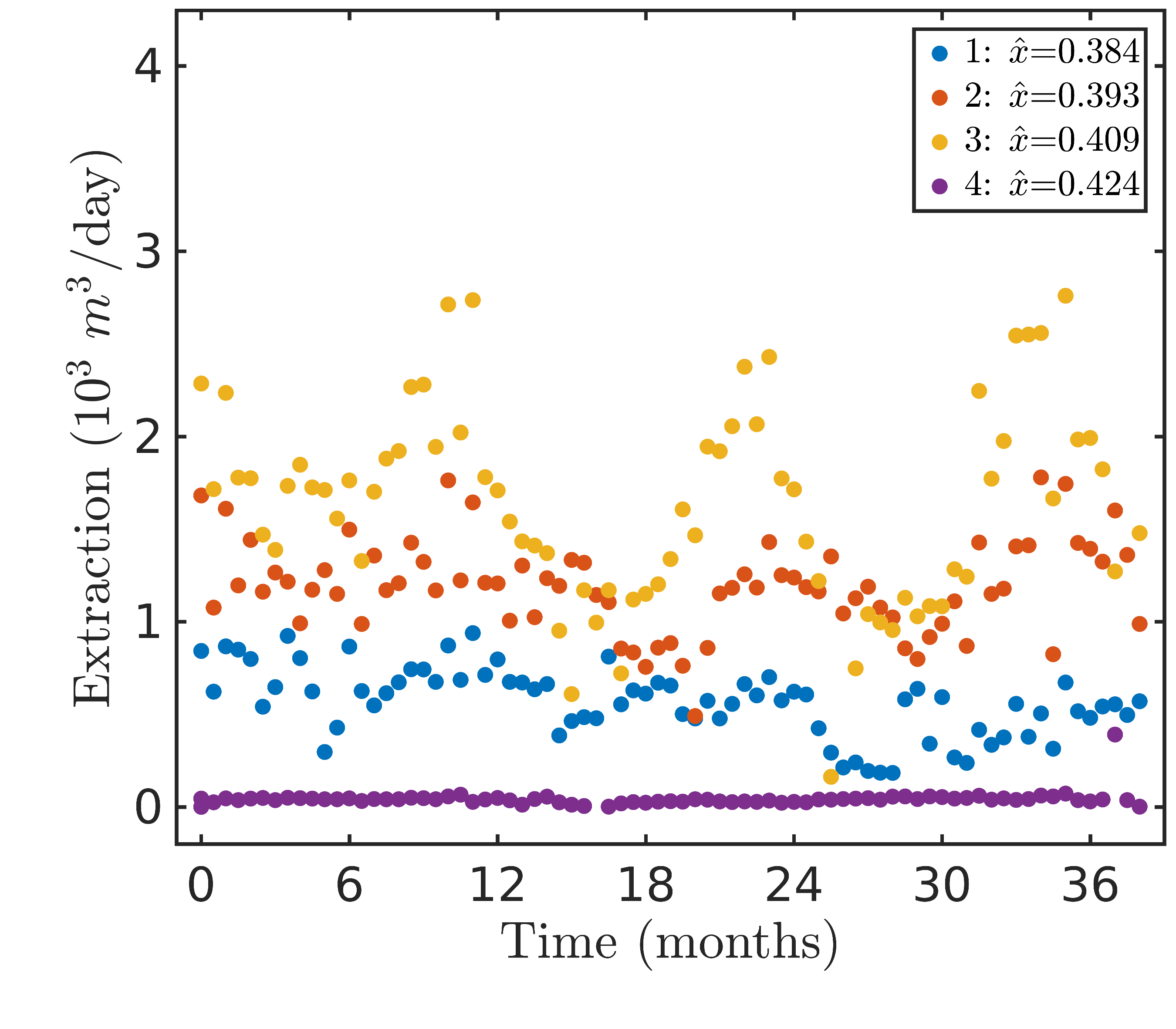}
\put(20,77){(c)}
\end{overpic}

\vspace{5mm}

\begin{overpic}[width=0.32\textwidth]{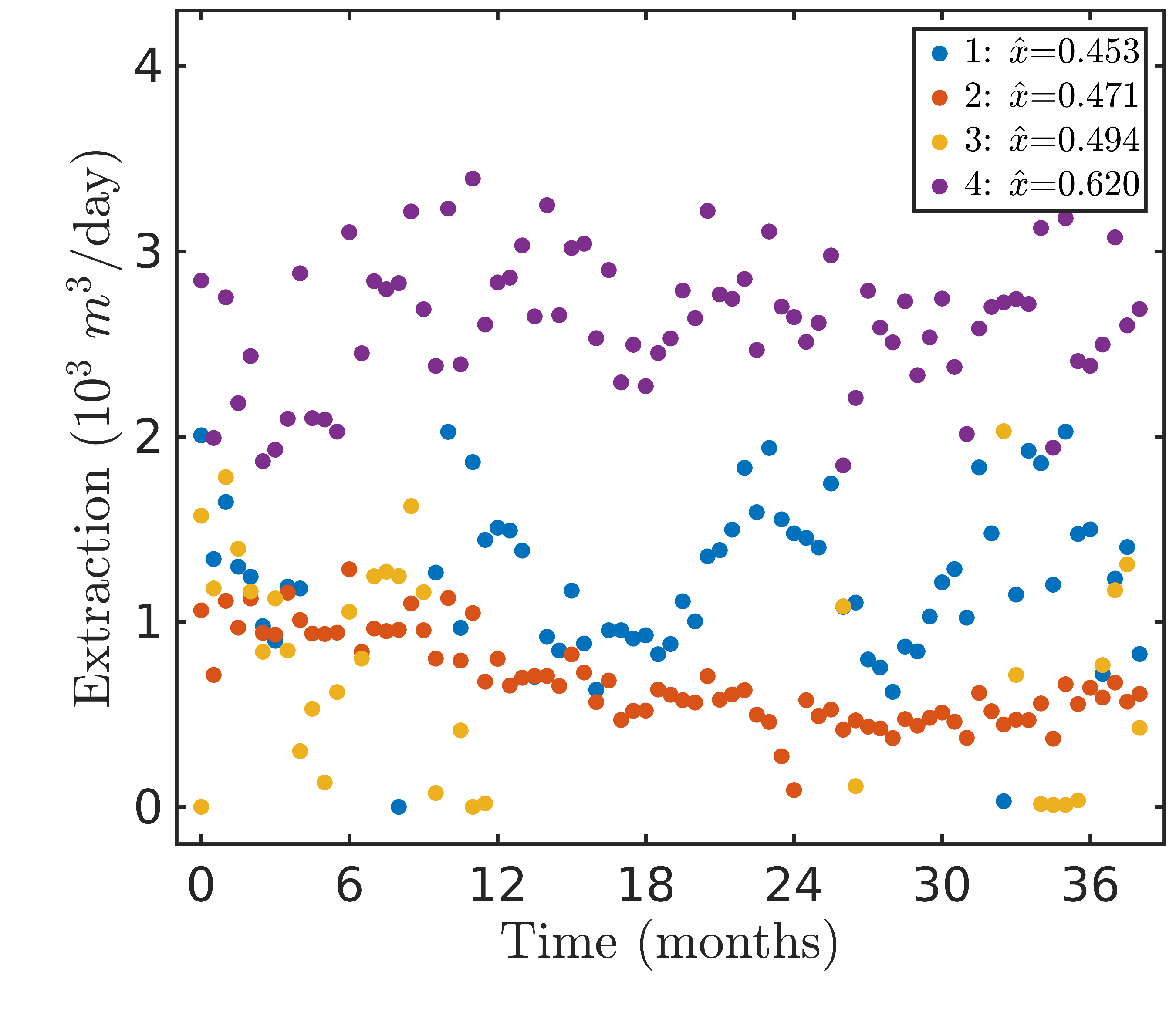}
\put(20,77){(d)}
\end{overpic}
\hfill
\begin{overpic}[width=0.32\textwidth]{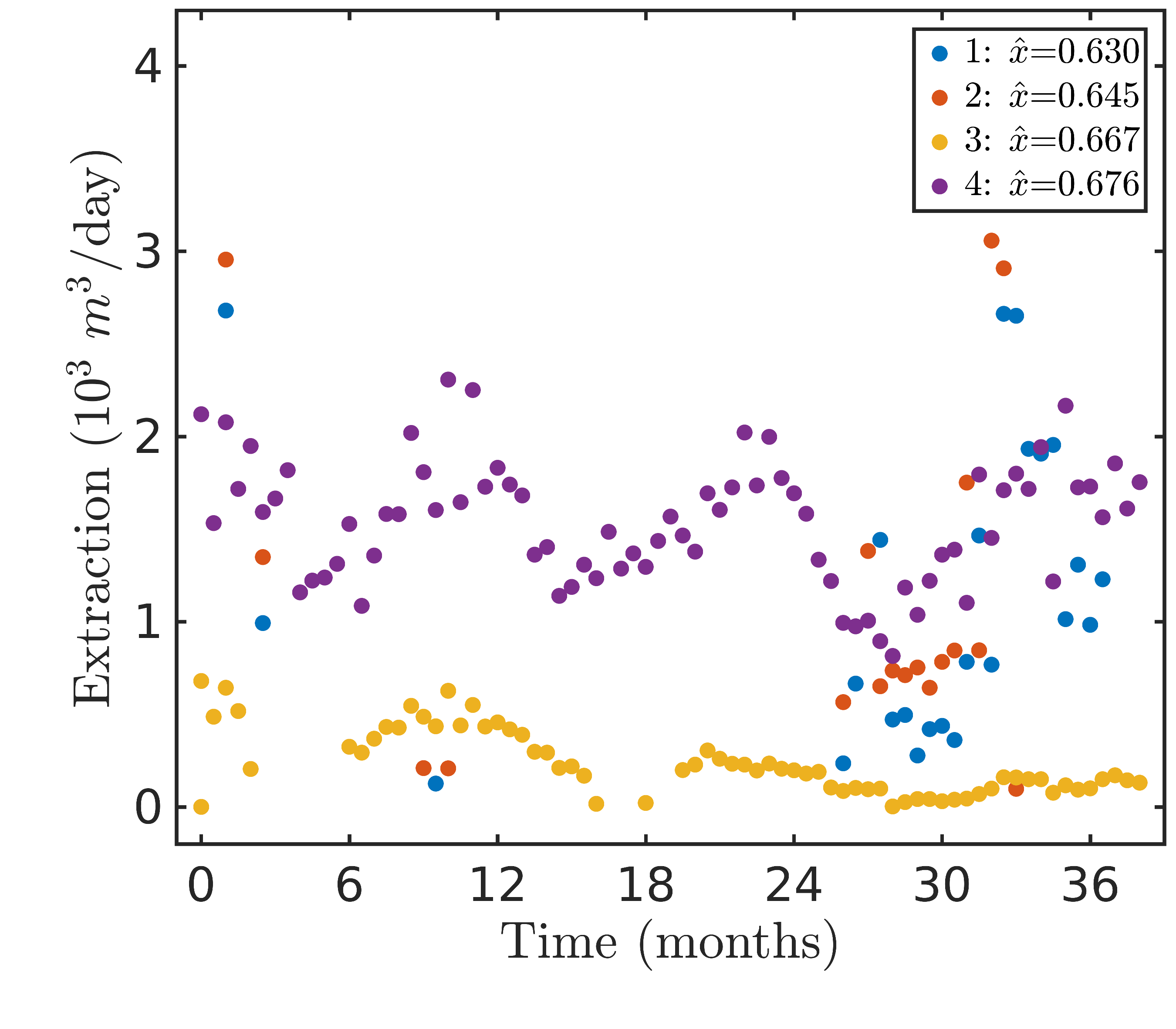}
\put(20,77){(e)}
\end{overpic}
\hfill
\begin{overpic}[width=0.32\textwidth]{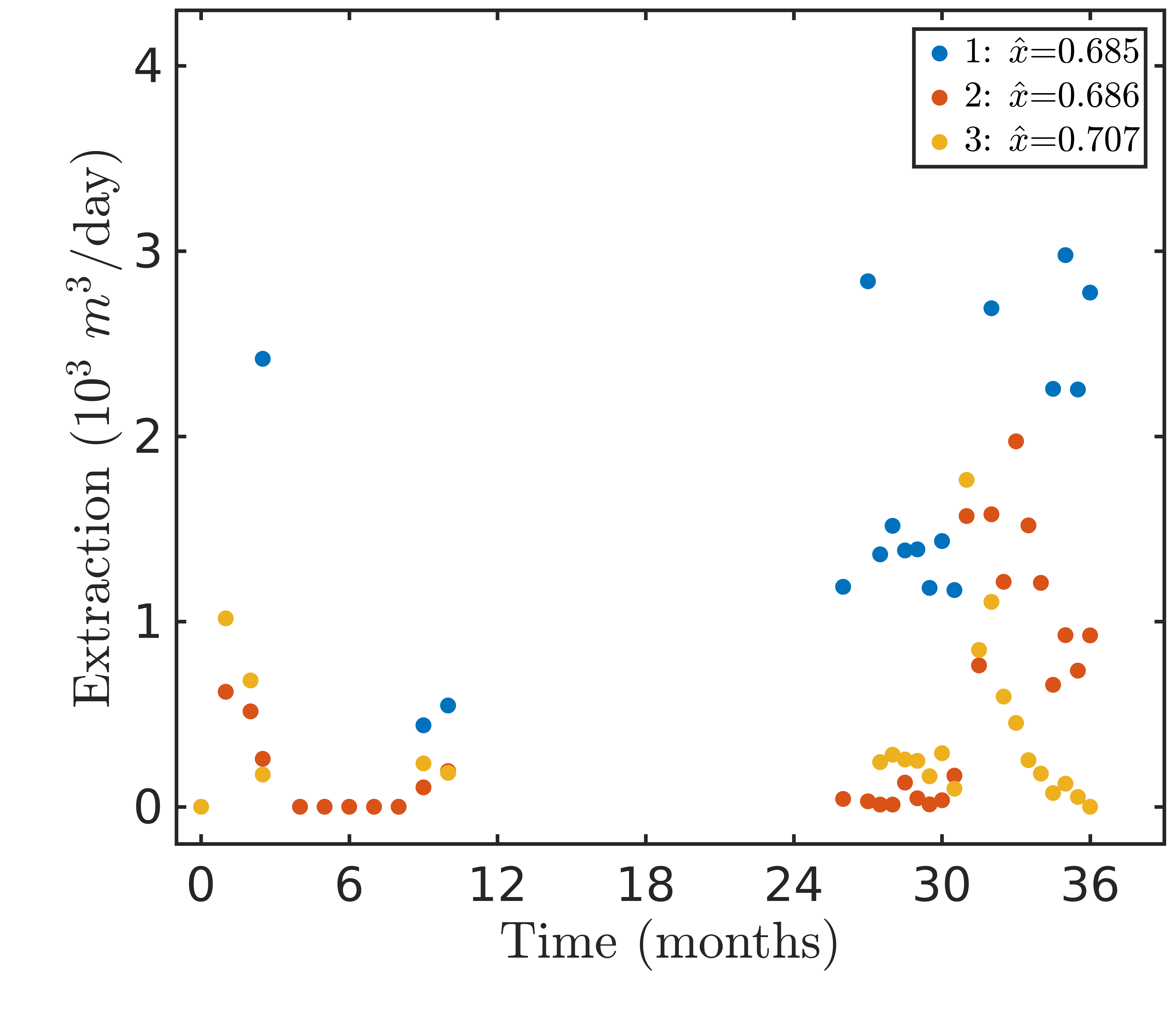}
\put(20,77){(f)}
\end{overpic}
\caption{ (a) Time series data (from October 2013 - November 2016) of the recharge and dam seepage rates. The location of the recharge sources $R_{1-4}$ are $\hat{x}= 0.04,\,0.26,\,0.51,\,0.68.$ 
(b-f) Transient extraction rates of the nineteen bore wells. The location of the extractions are mentioned in the corresponding legends.}
\label{fig:timeseries_data}
\end{figure}

\clearpage \newpage
\section{Numerical details for sea water intrusion \label{appB}}

The numerical computations are carried out using the finite-element-based commercial package COMSOL~v5.2~\cite{pryor2009multiphysics, li2009comsol}. The discretization of the variational form of the governing equations follows the standard Galerkin formalism \cite{hu2001direct}. To save computational cost, the Jacobian matrix is updated once in several iterations. Within each Newton iteration, the sparse linear system is solved by preconditioned Krylov methods \cite{elman2014finite} such as the generalized minimum residual (GMRES) method \cite{saad1986gmres, saad1993flexible} and the bi-conjugate gradient stabilized (BCGSTAB) method \cite{barrett1994templates}. The segregated solver \cite{pryor2009multiphysics} is used with a relative tolerance of 0.001. Two segregated steps decoupling the hydrodynamics and the director equations are set up.

The time-dependent solver is utilized with a relative tolerance of $10^{-4}$ along with the moving mesh method based on Arbitrary Lagrangian--Eulerian (ALE) formulation \cite{hu2001direct, donea1982arbitrary, anderson2004arbitrary}, which enables the moving boundaries without the need for mesh movement. This allows for mesh deformation with the translation of the physical boundary. Smoothing the deformation (because of the movement of the interior nodes) throughout the domain is achieved by solving the hyperelastic \cite{yamada1993arbitrary, curtis2013axisymmetric} smoothing partial differential equation with the boundary conditions specified. Second-order shape elements are used to mesh the geometry in the case of free deformation with particle movement. The spatial derivatives of the dependent variables are now transformed with respect to the new transformed coordinates. When the mesh deformation produces a low-element quality (defined as the ratio of the inscribed to circumscribed circles’ radii for each simplex element), where the minimum element quality is less than 0.3 (it is 0 for the degenerated element and 1 for the best possible element), remeshing of the entire domain is done and the simulation progresses. 

\begin{figure}[!htb]
\centering
\footnotesize
\begin{overpic}[width=1\textwidth]{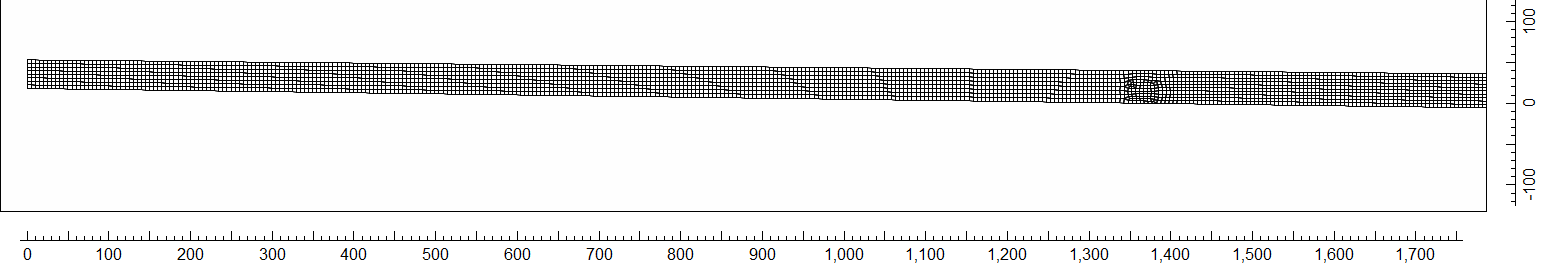}
\end{overpic}
\caption{
\rv{Geometry of the model domain, based on the dimensions and orientation shown in Fig.~\ref{fig:schem} which is used for the SUTRA simulation. The irregular meshing method is applied here.}
}
\label{fig:sutra-mesh}
\end{figure}

\rv{
Regarding the SUTRA simulation, the present situation is similar to the Henry problem. Initially a 2D rectangular box was created, representing the calculation domain, as shown in Fig.~\ref{fig:schem}, except we extend the length of the aquifer to be $2L$ so as to account for the downstream sea region, as described in Section~\ref{sec:sea_intru}. The fresh water is on the left side and the sea water is through the right boundary, similar to the configuration described in Fig.~\ref{fig:schem}. We have used an irregular mesh generation method \mbox{\cite{Cuthill1969}} with an average mesh size of 5m, as illustrated in Fig.~\ref{fig:sutra-mesh}. The transport scheme selected is ``solute using pressure" and the flow conditions are considered to be saturated. The boundary conditions as presented in the current work are incorporated in this simulation. For the physical properties of the aquifer, we have considered the hydraulic conductivity of the domain to be constant at 130 m/day and the porosity was set to 0.4. The ratio of the density of sea water to freshwater is considered to be 1.03 ($\beta = 0.03$). For the operating conditions we have used constant extraction rates equivalent to the values of October 2013 at all the 19 locations, and the recharge rates are switched off to clearly illustrate the effect of the sea water intrusion. The dam seepage rate is assumed to be constant at the rate of 500 $m^3$/day. The model is set up using the ModelMuse v3.10 GUI package \mbox{\cite{burnette2016situ}} which is the subsequently fed in to the SUTRA code.
First the simulation is run for steady state to generate the initial data and then it is ``restarted" using the result (of the steady state run) as an initial condition for the transient calculations over a 6 year period. The contour lines for the concentration are for $\beta = 0.005$ (a measure of the density ratio) plotted at different times, as shown in Fig.~\ref{fig:sea_water_intrusion_2}b.
}
\vspace{5mm}

\clearpage \newpage
\section{Contaminant transport \label{appC}}
In the event of an accidental spillage or seepage of chemical contamination in an aquifer, it is important to understand and forecast the impact of contamination. For simplicity, we restrict our attention to the case where the contamination is uniform in the vertical direction $\hat{z}$. Therefore, the contaminant undergoes both horizontal advection by the fluid velocity $\hat{u}$ (see equation (15)), and molecular diffusion. In practice, usually the contamination volume is small relative to the volume of the aquifer, so that it does not affect the flow field significantly, and we proceed with this assumption. In this way, the transport of contaminant (species) concentration $c$ (in the dilute range) can be modelled with an one-dimensional advection-diffusion equation. Here we define the non-dimensional concentration $\hat{c}=c/c_0$, where $c_0$ is a typical concentration value, such that
\begin{equation} \label{eq:cont_transp}
\frac{\partial \hat{c}}{\partial \hat{t}} + \hat{u} \frac{\partial \hat{c}}{\partial \hat{x}} = \frac{1}{{\rm Pe}_c} \frac{\partial^2 \hat{c}}{\partial \hat{x}^2} + G(\hat{x},\hat{t}),
\end{equation}
where the Peclet number Pe$_c = KH/D_c$ is given in terms of the hindered diffusivity of the contaminant in the porous medium, $D_c \approx \phi D_{c0}$ (where $D_{c0}$ is the bulk diffusivity). \rvs{Here we also do not consider the hydrodynamic dispersion effects, so effectively $D_c \not\approx f(u)$.} We also assume that the spillage occurs at an isolated location for a finite time $\hat{x}_c$ for $0\leq\hat{t}=[\leq \hat{t}_c]$ (``pulse" spillage). In equation \eqref{eq:cont_transp}, $G$ represents the spillage and can be approximated by a Dirac-delta function $\delta$, such that
\begin{equation} \label{G_fun}
G(\hat{x},\hat{t})=
	\begin{cases}
	\delta \left( \hat{x} - \hat{x}_c \right) &~\mathrm{for}~\hat{t} \le \hat{t}_c, \\
    0 &~\mathrm{for}~\hat{t} > \hat{t}_c.
	\end{cases}
\end{equation}
We consider a very diffusive solute with $D_{c_0}= 2\times 10^{-9} \un{m^2/s}$, which leads to a Peclet number Pe$_c \approx 2 \times 10^8$. Even with a very diffusive solute, we still find Pe$_c \gg 1$, and hence the contaminant transport is advection dominated.

Therefore, it is insightful to analyse the worst case scenario, when the spill happens at the dam ($\hat{x}_c = 0$). The boundary condition at the far end of the aquifer ($\hat{x} = 1$) is the outlet condition that there is no change in the spatial concentration there, that is
\begin{equation}
\frac{\partial \hat{c}}{\partial \hat{x}}=0 \quad \mathrm{at} \quad \hat{x}=1.\label{far_field_con}
\end{equation}
The boundary condition at the dam $(\hat{x}=0)$ is also that no concentration flux exists there, since we do not expect the contaminant to diffuse across the dam boundary.
Furthermore, we assume that at the time of the spillage, there is no contaminant anywhere in the aquifer, so we have the initial condition
\begin{equation} \label{c_c_ics}
\hat{c}=0 \quad \mathrm{at}\quad \hat{t}=0.
\end{equation}

We solve equations (\ref{eq:cont_transp})$-$(\ref{c_c_ics}) for the concentration of a chemical contaminant $\hat{c}$, given the spillage function (\ref{G_fun}), where the advection velocity $\hat{u}$ is calculated using the Dupuit-Forchheimer model (Equations (\ref{govh})$-$(\ref{nondimbc2})) with recharge, extraction and dam seepage rates taken from the Germasogeia aquifer data with Gaussian distributions for the recharge sources (as described in Section 2). The permeability of the aquifer is considered constant, $\hat{k}=1$.
Since the contaminant travels downstream due to fluid advection, the spillage affects more if it occurs upstream. 

\begin{figure}[!htbp]
\centering
\footnotesize
\begin{overpic}[width=0.44\textwidth]{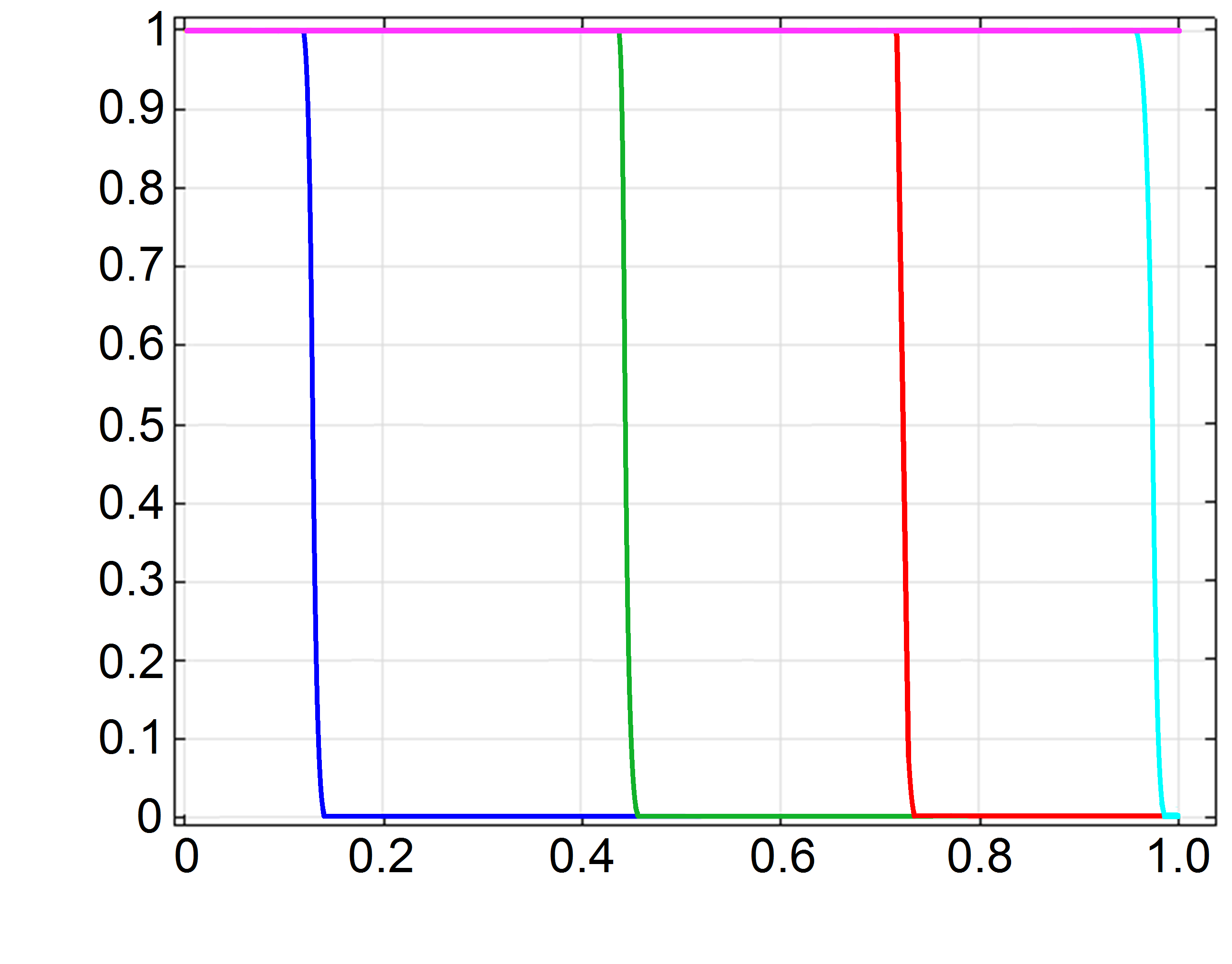}
\put(-10,70){\normalsize (a)}
\put(0,10){\rotatebox{90}{Contaminant concentration $(\hat{c})$}}
\put(25,0){\rotatebox{0}{Scaled distance from dam $(\hat{x})$}}
\put(27,30){\rotatebox{90}{$\hat{t}=0.025$}}
\put(44,30){\rotatebox{90}{$\hat{t}=0.125$}}
\put(67,30){\rotatebox{90}{$\hat{t}=0.22$}}
\put(87,30){\rotatebox{90}{$\hat{t}=0.32$}}
\put(30,70){\rotatebox{0}{$\hat{t}=0.37$}}
\end{overpic}
\hfill
\begin{overpic}[width=0.46\textwidth]{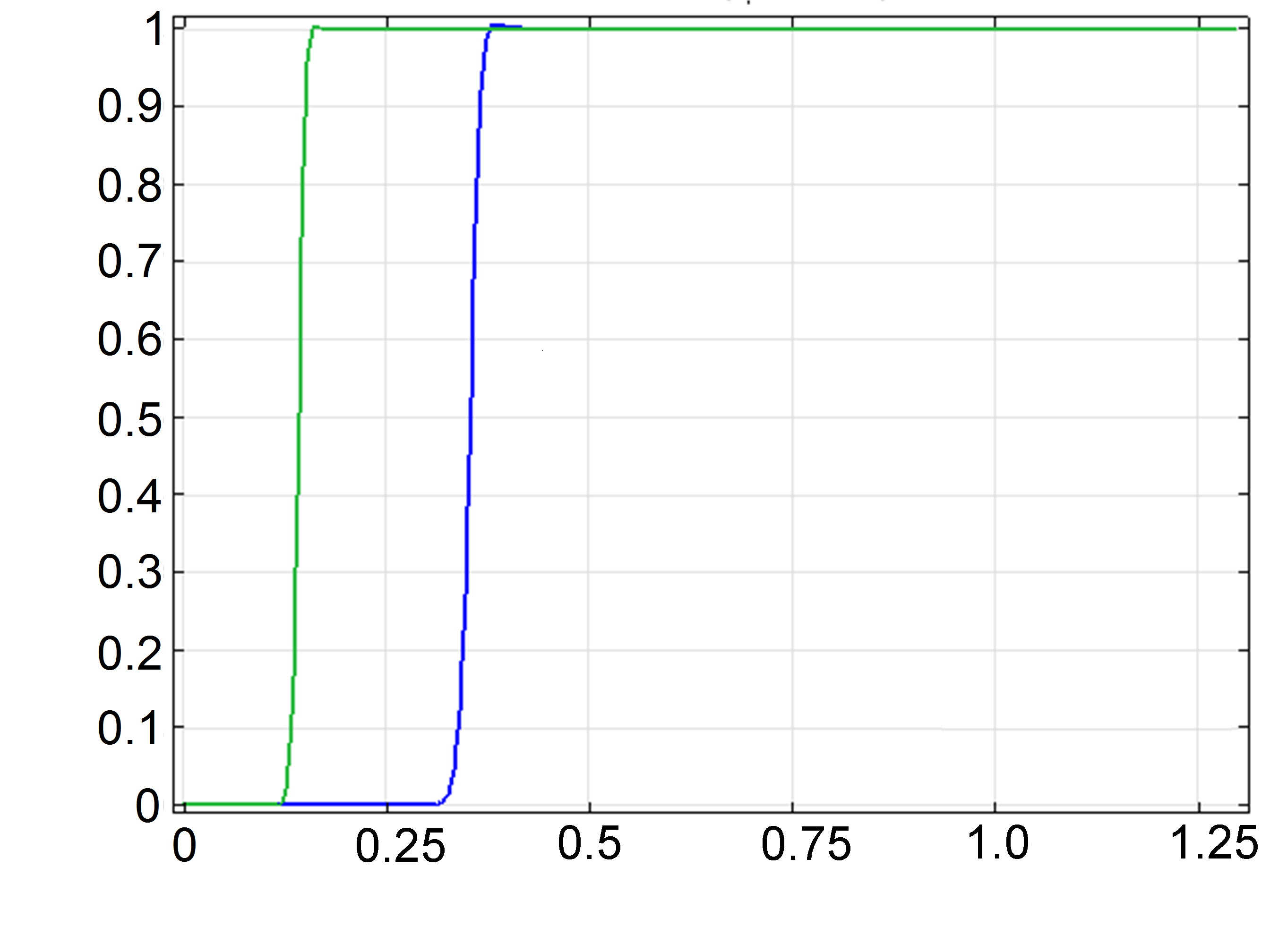}
\put(-10,70){\normalsize (b)}
\put(40,0){Scaled time $(\hat{t})$}
\put(0,10){\rotatebox{90}{Contaminant concentration $(\hat{c})$}}
\put(25,35){\rotatebox{90}{$\hat{x}=0.5$}}
\put(38,25){\rotatebox{90}{$\hat{x}=1$}}
\end{overpic}
\caption{Plots of the contaminant concentration profile ($\hat{c}$) for a continuous input spillage ($\hat{t}_c \rightarrow \infty $) at the dam, $\hat{x} = 0$, upstream of the aquifer at (a) different time instants, $\hat{t}$ and (b) evolving with time at different spatial locations ($\hat{x}$). The conductivity of the medium is considered constant (130 m/day) and Gaussian distributions for the recharge inputs have been fitted, as previously. The dam seepage, recharge and extraction rates are from field data between Oct 2013--Nov 2016, as used in Fig. \ref{fig:matching}.
}
\label{fig:contaminant_const_input}
\end{figure}
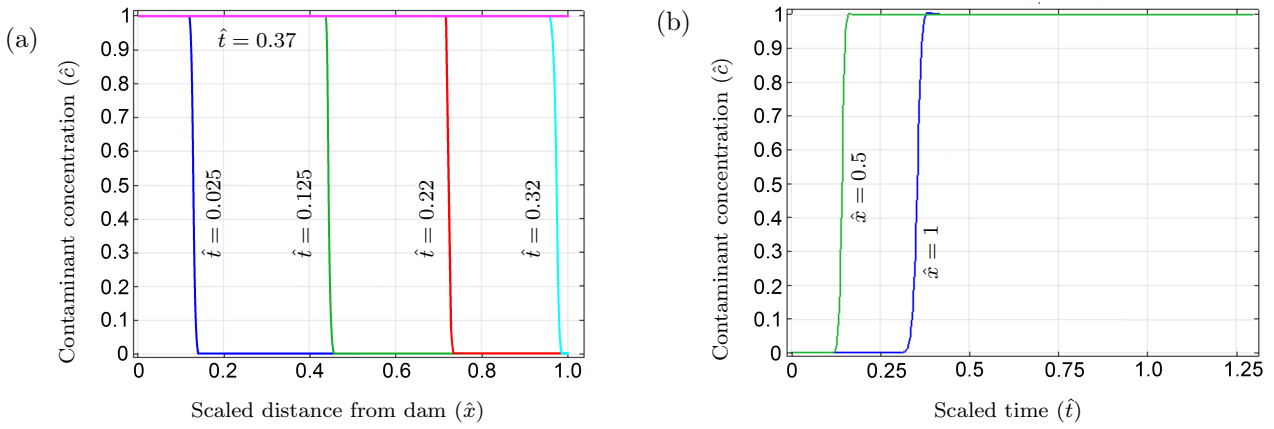

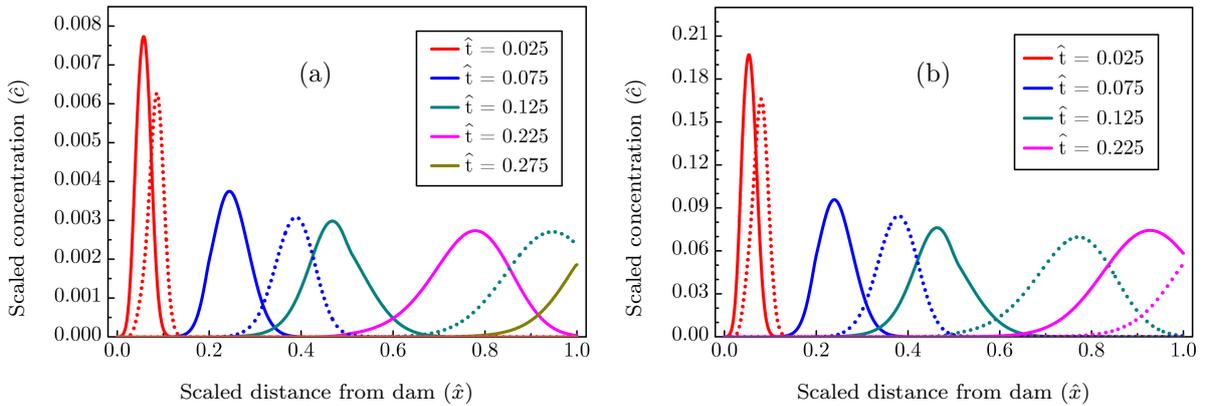
\begin{figure}[!htbp]
\footnotesize
\centering
\begin{overpic}[width=0.49\textwidth]{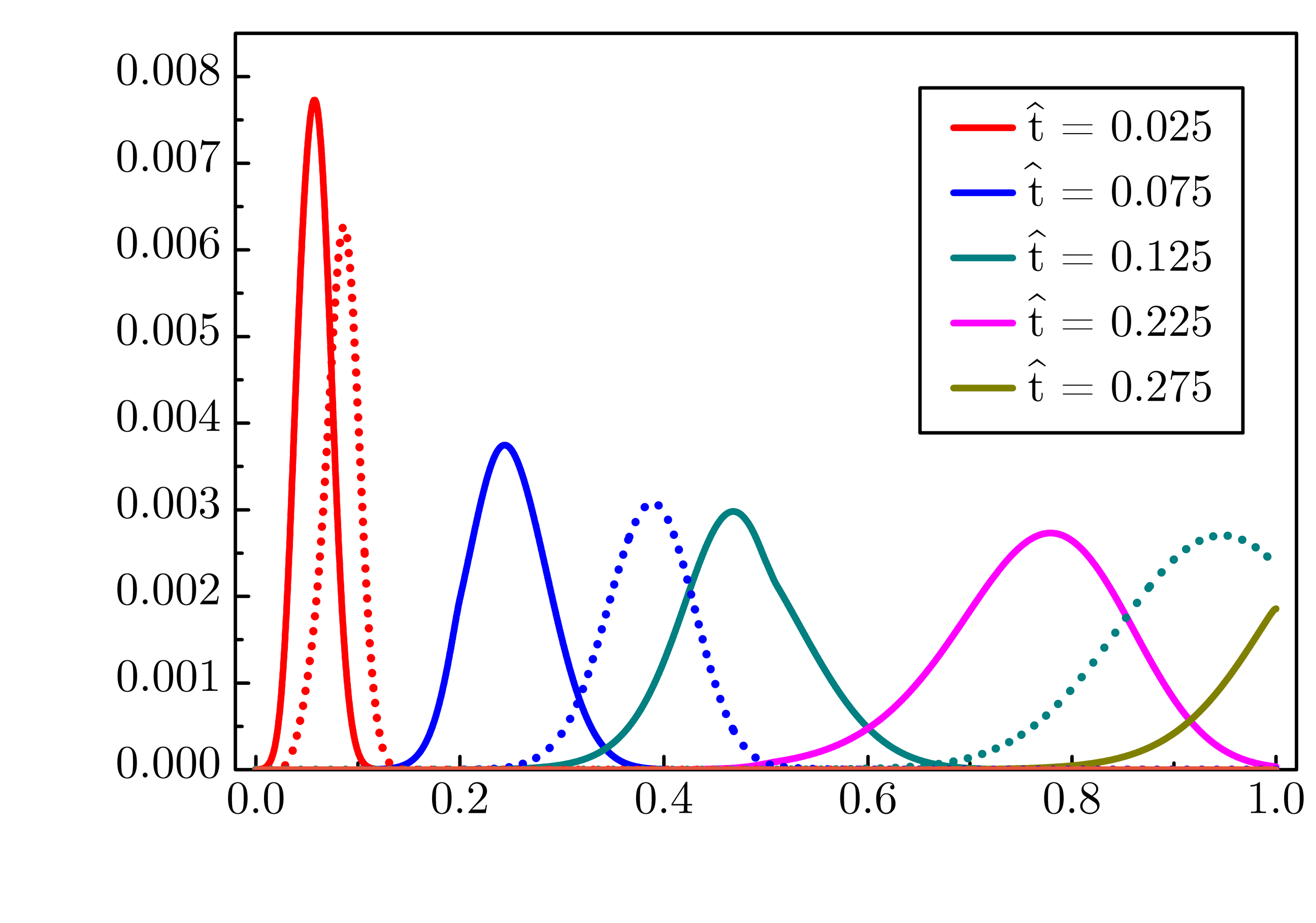}
\put(50,55){\normalsize (a)}
\put(30,1){Scaled distance from dam $(\hat{x})$}
\put(1,15){\rotatebox{90}{Scaled concentration $(\hat{c})$}}
\end{overpic}
\hfill
\begin{overpic}[width=0.49\textwidth]{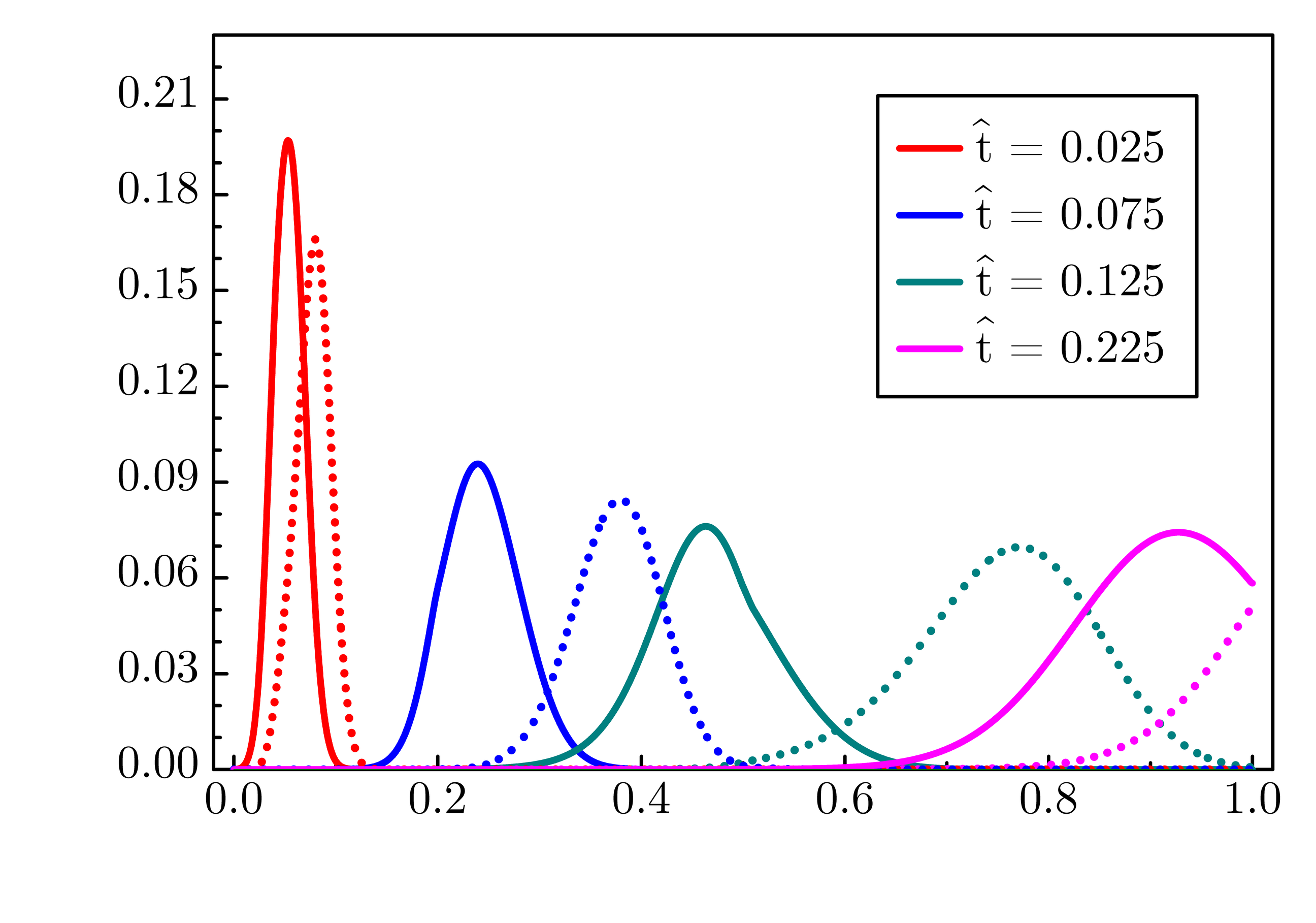}
\put(50,55){\normalsize (b)}
\put(30,1){Scaled distance from dam $(\hat{x})$}
\put(1,15){\rotatebox{90}{Scaled concentration $(\hat{c})$}}
\end{overpic}
\caption{Plots of the contaminant concentration profile ($\hat{c}$) in the aquifer for a pulse input (spillage) near the dam, $\hat{x} = 0$ which is spilled for (a) 1 day ($\hat{t}_c=3.4\times10^{-4}$) and (b) 30 days ($\hat{t}_c=0.01$) at constant concentration ($\hat{c} = 1$). The permeability of the medium has been considered constant throughout the $\hat{x}$ domain (130 m/day) and the Gaussian distribution of the recharge input is considered. The dam seepage, recharge and extraction rates are the same for the study period of the Germasogeia aquifer (Oct 2013--Nov 2016) as used in Fig.~\ref{fig:matching}. For the dotted curves, we have considered doubled recharge rates.}
\label{fig:contamination_pulse_input}
\end{figure}

We also consider different spillage durations $\hat{t}_c$ . In the first case we choose a continuous (uninterrupted) spillage ($\hat{t}_c \rightarrow \infty$), in the second case a spillage time at $\hat{t}_c = 3.4\times 10^{-4}$ (1 day), and in the third case at $\hat{t}_c = 0.01$ (30 days). 
The second and third cases correspond to situations in which the spillage is detected and contained successfully. In Figs. \ref{fig:contaminant_const_input} and \ref{fig:contamination_pulse_input}, we show how the contaminant is transported across the aquifer as time progresses for the three cases considered.

In the first case  (Fig. \ref{fig:contaminant_const_input}) we see that the contaminant propagation front is almost vertical, which is explained by the fact that the Peclet number is very large. This suggests that a monitoring device would detect a very sudden change in contaminant concentration levels (zero to unity within a couple of weeks  as seen in Fig. \ref{fig:contaminant_const_input}b). We can see that if the contamination spillage is undetected and unhindered, the aquifer gets completely polluted within 3 years ($\hat{t} \approx 0.37$). 

In the second and third cases, where the contaminant spillage is detected, the spillage source is stopped in a finite time. The contaminant, which enters the aquifer, gets advected and eventually the concentration levels throughout the aquifer fall to an acceptable level $(\varepsilon_c)$. 
The concentration profile at different times is displayed for the second case in Fig. \ref{fig:contamination_pulse_input}(a) and for the third case in Fig. \ref{fig:contamination_pulse_input}(b). 
In each case, the contaminant transportation behaviour is similar, with the concentration isolated to a pulse which spreads out over time. In the case of the longer spillage, the magnitude of the concentration is higher overall. The results also provide information about the time dependent position of the contaminant and one can infer the region of safety ($\hat{c} < \varepsilon_c$).

To illustrate the effect of recharge on contaminant transport, we display the concentration profiles for double recharge rates as dotted lines in Fig.~\ref{fig:contamination_pulse_input}. We clearly see that by increasing recharge, the contaminant is flushed more quickly. Furthermore, in the case of the briefer spillage, increasing the recharge has a greater flushing effect than for the longer spillage (note the profiles for $\hat{t}=0.0125$ in Figs.~\ref{fig:contamination_pulse_input}a and b).


\small
\bibliographystyle{unsrt}
\bibliography{bib_file}

\begin{thebibliography}{10}

\bibitem{theodossiou2004application}
N.~P. Theodossiou.
\newblock Application of non-linear simulation and optimisation models in
  groundwater aquifer management.
\newblock {\em Water Resources Management}, 18(2):125--141, 2004.

\bibitem{wanakule1986optimal}
N.~Wanakule, L.~W. Mays, and L.~S. Lasdon.
\newblock Optimal management of large-scale aquifers: Methodology and
  applications.
\newblock {\em Water Resources Research}, 22(4):447--465, 1986.

\bibitem{hallaji1996optimal}
K.~Hallaji and H.~Yazicigil.
\newblock {Optimal management of a coastal aquifer in southern Turkey}.
\newblock {\em Journal of Water Resources Planning and Management},
  122(4):233--244, 1996.

\bibitem{sedki2011simulation}
A.~Sedki and D.~Ouazar.
\newblock Simulation-optimization modeling for sustainable groundwater
  development: a moroccan coastal aquifer case study.
\newblock {\em Water Resources Management}, 25(11):2855--2875, 2011.

\bibitem{willis1988planning}
R.~Willis and B.~A. Finney.
\newblock Planning model for optimal control of saltwater intrusion.
\newblock {\em Journal of Water Resources Planning and Management},
  114(2):163--178, 1988.

\bibitem{mantoglou2003pumping}
A.~Mantoglou.
\newblock Pumping management of coastal aquifers using analytical models of
  saltwater intrusion.
\newblock {\em Water Resources Research}, 39(12), 2003.

\bibitem{basha2005theoretical}
H.~A Basha and S.~F. Maalouf.
\newblock Theoretical and conceptual models of subsurface hillslope flows.
\newblock {\em Water Resources Research}, 41(7), 2005.

\bibitem{castro2012steady}
O.~Castro-Orgaz and J.~V. Gir{\'a}ldez.
\newblock Steady-state water table height estimations with an improved
  pseudo-two-dimensional dupuit-forchheimer type model.
\newblock {\em Journal of Hydrology}, 438:194--202, 2012.

\bibitem{van2017interface}
C.~J. van Duijn and R.~J. Schotting.
\newblock {The Interface Between Fresh and Salt Groundwater in Horizontal
  Aquifers: The Dupuit-Forchheimer Approximation Revisited}.
\newblock {\em Transport in Porous Media}, 117(3):481--505, 2017.

\bibitem{koussis2012analytical}
A.~D. Koussis, K.~Mazi, and G.~Destouni.
\newblock Analytical single-potential, sharp-interface solutions for regional
  seawater intrusion in sloping unconfined coastal aquifers, with pumping and
  recharge.
\newblock {\em Journal of Hydrology}, 416:1--11, 2012.

\bibitem{koussis2015correction}
A.~D. Koussis, K.~Mazi, F.~Riou, and G.~Destouni.
\newblock {A correction for Dupuit-Forchheimer interface flow models of
  seawater intrusion in unconfined coastal aquifers}.
\newblock {\em Journal of Hydrology}, 525:277--285, 2015.

\bibitem{wooding1966groundwater}
R.~A. Wooding and T.~G. Chapman.
\newblock {Groundwater flow over a sloping impermeable layer: 1. Application of
  the Dupuit-Forchheimer assumption}.
\newblock {\em Journal of Geophysical Research}, 71(12):2895--2902, 1966.

\bibitem{jones1987optimal}
L.~Jones, R.~Willis, and W.~W.~G. Yeh.
\newblock Optimal control of nonlinear groundwater hydraulics using
  differential dynamic programming.
\newblock {\em Water Resources Research}, 23(11):2097--2106, 1987.

\bibitem{makinde1989optimal}
B.~A Makinde-Odusola and M.~A. Mari{\~n}o.
\newblock Optimal control of groundwater by the feedback method of control.
\newblock {\em Water Resources Research}, 25(6):1341--1352, 1989.

\bibitem{cheng2000pumping}
A.~H.-D. Cheng, D.~Halhal, A.~Naji, and D.~Ouazar.
\newblock Pumping optimization in saltwater-intruded coastal aquifers.
\newblock {\em Water Resources Research}, 36(8):2155--2165, 2000.

\bibitem{casola1986optimal}
W.~H. Casola, R.~Narayanan, C.~Duffy, and A.~B. Bishop.
\newblock Optimal control model for groundwater management.
\newblock {\em Journal of Water Resources Planning and Management},
  112(2):183--197, 1986.

\bibitem{shamir1984optimal}
U.~Shamir, J.~Bear, and A.~Gamliel.
\newblock Optimal annual operation of a coastal aquifer.
\newblock {\em Water Resources Research}, 20(4):435--444, 1984.

\bibitem{strack1976single}
O.~D.~L. Strack.
\newblock A single-potential solution for regional interface problems in
  coastal aquifers.
\newblock {\em Water Resources Research}, 12(6):1165--1174, 1976.

\bibitem{badon1889nota}
W.~Badon-Ghyben.
\newblock Nota in verband met de voorgenomen putboring nabil amsterdam.
\newblock {\em Tijdschr. k. inst. ing., the Hague}, 27:1888--1889, 1889.

\bibitem{herzberg1901wasserversorgung}
A.~Herzberg.
\newblock Die wasserversorgung einiger nordseebader, munich.
\newblock {\em Jour. Gasbeleuchiung und Wasserver-sorgung}, 44:815--819, 1901.

\bibitem{de1981variational}
G.~De~Josselin De~Jong.
\newblock A variational fallacy.
\newblock {\em G{\'e}otechnique}, 31(2), 1981.

\bibitem{charmonman1965solution}
S.~Charmonman.
\newblock A solution of the pattern of fresh-water flow in an unconfined
  coastal aquifer.
\newblock {\em Journal of Geophysical Research}, 70(12):2813--2819, 1965.

\bibitem{henry1959salt}
H.~R. Henry.
\newblock Salt intrusion into fresh-water aquifers.
\newblock {\em Journal of Geophysical Research}, 64(11):1911--1919, 1959.

\bibitem{rumer1968salt}
R.~R. Rumer and J.~C. Shiau.
\newblock Salt water interface in a layered coastal aquifer.
\newblock {\em Water Resources Research}, 4(6):1235--1247, 1968.

\bibitem{henry1964effects}
H.~R. Henry.
\newblock Effects of dispersion on salt encroachment in coastal aquifers, in
  seawater in coastal aquifers.
\newblock {\em US Geological Survey, Water Supply Paper}, 1613:C70--C80, 1964.

\bibitem{bolster2007analytical}
D.~T. Bolster, D.~M. Tartakovsky, and M.~Dentz.
\newblock Analytical models of contaminant transport in coastal aquifers.
\newblock {\em Advances in Water Resources}, 30(9):1962--1972, 2007.

\bibitem{croucher1995henry}
A.~E. Croucher and M.~J. O'Sullivan.
\newblock The henry problem for saltwater intrusion.
\newblock {\em Water Resources Research}, 31(7):1809--1814, 1995.

\bibitem{simpson2004improving}
M.~J. Simpson and T.~P. Clement.
\newblock Improving the worthiness of the henry problem as a benchmark for
  density-dependent groundwater flow models.
\newblock {\em Water Resources Research}, 40(1), 2004.

\bibitem{dentz2006variable}
M.~Dentz, D.~M. Tartakovsky, E.~Abarca, A.~Guadagnini, X.~Sanchez-Vila, and
  J.~Carrera.
\newblock Variable-density flow in porous media.
\newblock {\em Journal of Fluid Mechanics}, 561:209--235, 2006.

\bibitem{bear2010modeling}
J.~Bear and A.~H.~D. Cheng.
\newblock {\em Modeling groundwater flow and contaminant transport}, volume~23.
\newblock Springer Science \& Business Media, 2010.

\bibitem{zheng2002applied}
C.~Zheng, G.~D. Bennett, et~al.
\newblock {\em Applied contaminant transport modeling}, volume~2.
\newblock Wiley-Interscience New York, 2002.

\bibitem{furman2008modeling}
A.~Furman.
\newblock Modeling coupled surface--subsurface flow processes: a review.
\newblock {\em Vadose Zone Journal}, 7(2):741--756, 2008.

\bibitem{joodi2010development}
A.~S. Joodi, S.~Sizaret, S.~Binet, A.~Bruand, P.~Alberic, and M.~Lepiller.
\newblock Development of a darcy-brinkman model to simulate water flow and
  tracer transport in a heterogeneous karstic aquifer (val d’orl{\'e}ans,
  france).
\newblock {\em Hydrogeology journal}, 18(2):295--309, 2010.

\bibitem{chen2010asymptotic}
N~Chen, M~Gunzburger, and X~Wang.
\newblock Asymptotic analysis of the differences between the stokes--darcy
  system with different interface conditions and the stokes--brinkman system.
\newblock {\em Journal of Mathematical Analysis and Applications},
  368(2):658--676, 2010.

\bibitem{neale1974practical}
G.~Neale and W.~Nader.
\newblock Practical significance of brinkman's extension of darcy's law:
  coupled parallel flows within a channel and a bounding porous medium.
\newblock {\em The Canadian Journal of Chemical Engineering}, 52(4):475--478,
  1974.

\bibitem{llopis2014discussion}
Carlos Llopis-Albert and David Pulido-Velazquez.
\newblock Discussion about the validity of sharp-interface models to deal with
  seawater intrusion in coastal aquifers.
\newblock {\em Hydrological Processes}, 28(10):3642--3654, 2014.

\bibitem{nocedal2006numerical}
J.~Nocedal and S.~J. Wright.
\newblock Numerical optimization, second edition, 2006.

\bibitem{wachter2006implementation}
A.~W{\"a}chter and L.~T. Biegler.
\newblock On the implementation of an interior-point filter line-search
  algorithm for large-scale nonlinear programming.
\newblock {\em Mathematical Programming}, 106(1):25--57, 2006.

\bibitem{dunning2017jump}
I.~Dunning, J.~Huchette, and M.~Lubin.
\newblock Jump: A modeling language for mathematical optimization.
\newblock {\em SIAM Review}, 59(2):295--320, 2017.

\bibitem{bezanson2017julia}
J.~Bezanson, A.~Edelman, S.~Karpinski, and V.~B. Shah.
\newblock Julia: A fresh approach to numerical computing.
\newblock {\em SIAM Review}, 59(1):65--98, 2017.

\bibitem{cussler2009diffusion}
E.~L. Cussler.
\newblock {\em Diffusion: Mass Transfer in Fluid Systems}.
\newblock Cambridge Series in Chemical Engineering. Cambridge University Press,
  2009.

\bibitem{voss2002sutra}
CI~Voss and AM~Provost.
\newblock Sutra, a model for saturated--unsaturated, variable-density
  ground-water flow with solute or energy transport. water-resour. invest. rep.
  2002-4231. usgs, reston, va.
\newblock {\em SUTRA, a model for saturated--unsaturated, variable-density
  ground-water flow with solute or energy transport. Water-Resour. Invest. Rep.
  2002-4231. USGS, Reston, VA.}, 2002.

\bibitem{pryor2009multiphysics}
R.~W. Pryor.
\newblock {\em Multiphysics modeling using {COMSOL}: a first principles
  approach}.
\newblock Jones \& Bartlett Publishers, 2009.

\bibitem{li2009comsol}
Q.~Li, K.~Ito, Z.~Wu, C.~S. Lowry, I.~I. Loheide, and P.~Steven.
\newblock {COMSOL} multiphysics: A novel approach to ground water modeling.
\newblock {\em Ground Water}, 47(4):480--487, 2009.

\bibitem{hu2001direct}
H.~H. Hu, N.~A. Patankar, and M.~Y. Zhu.
\newblock Direct numerical simulations of fluid--solid systems using the
  arbitrary {L}agrangian--{E}ulerian technique.
\newblock {\em J.~Comp.~Phys.}, 169(2):427--462, 2001.

\bibitem{elman2014finite}
H.~C. Elman, D.~J. Silvester, and A.~J. Wathen.
\newblock {\em Finite elements and fast iterative solvers: with applications in
  incompressible fluid dynamics}.
\newblock Oxford University Press, 2014.

\bibitem{saad1986gmres}
Y.~Saad and M.~H. Schultz.
\newblock Gmres: A generalized minimal residual algorithm for solving
  nonsymmetric linear systems.
\newblock {\em SIAM J.~Scientific and Statistical Computing}, 7(3):856--869,
  1986.

\bibitem{saad1993flexible}
Y.~Saad.
\newblock A flexible inner-outer preconditioned {GMRES} algorithm.
\newblock {\em SIAM J.~Scientific Computing}, 14(2):461--469, 1993.

\bibitem{barrett1994templates}
R.~Barrett, M.~W. Berry, T.~F. Chan, J.~Demmel, J.~Donato, J.~Dongarra,
  V.~Eijkhout, R.~Pozo, C.~Romine, and H.~Van~der Vorst.
\newblock {\em Templates for the solution of linear systems: building blocks
  for iterative methods}, volume~43.
\newblock SIAM, 1994.

\bibitem{donea1982arbitrary}
J.~Donea, S.~Giuliani, and J.-P. Halleux.
\newblock An arbitrary {L}agrangian-{E}ulerian finite element method for
  transient dynamic fluid--structure interactions.
\newblock {\em Computer Methods in Applied Mechanics and Engineering},
  33(1-3):689--723, 1982.

\bibitem{anderson2004arbitrary}
R.~W. Anderson, N.~S. Elliott, and R.~B. Pember.
\newblock An arbitrary {L}agrangian--{E}ulerian method with adaptive mesh
  refinement for the solution of the euler equations.
\newblock {\em J.~Comp.~Phys.}, 199(2):598--617, 2004.

\bibitem{yamada1993arbitrary}
T.~Yamada and F.~Kikuchi.
\newblock An arbitrary {L}agrangian-{E}ulerian finite element method for
  incompressible hyperelasticity.
\newblock {\em Computer Methods in Applied Mechanics and Engineering},
  102(2):149--177, 1993.

\bibitem{curtis2013axisymmetric}
C.~W. Curtis and M.~L. Calvisi.
\newblock Axisymmetric model of an intracranial saccular aneurysm: Theory and
  computation.
\newblock In {\em ASME 2013 International Mechanical Engineering Congress and
  Exposition}, pages V009T10A054--V009T10A054. American Society of Mechanical
  Engineers, 2013.

\bibitem{Cuthill1969}
Elizabeth Cuthill and James McKee.
\newblock Reducing the bandwidth of sparse symmetric matrices.
\newblock In {\em Proceedings of the 1969 24th national conference}, pages
  157--172. ACM, 1969.

\bibitem{burnette2016situ}
Matthew~C Burnette, David~P Genereux, and Fran{\c{c}}ois Birgand.
\newblock In-situ falling-head test for hydraulic conductivity: evaluation in
  layered sediments of an analysis derived for homogenous sediments.
\newblock {\em Journal of Hydrology}, 539:319--329, 2016.

\end{thebibliography}

\end{document}